\documentclass[11pt, oneside]{article}   	
\usepackage[margin=1in]{geometry}               
\usepackage{graphicx}				
										
\usepackage{amssymb}
\usepackage{physics}
\usepackage{amsmath}
\usepackage[dvips]{epsfig}

\usepackage{amsfonts}
\usepackage{subfig}

\usepackage{mathrsfs}

\usepackage{bm}

\usepackage[title]{appendix}

\usepackage{cite}

\usepackage{authblk}

\def\cchi{\raise2pt\hbox{$\chi$}} 

\newcommand\bsdot{\ensuremath{\boldsymbol{.}}}

\title{\bf{The effect of the width of the incident pulse to the dielectric transition layer in the scattering of an electromagnetic pulse - a quantum lattice algorithm simulation}
}

\author{George Vahala}
\affil{Department of Physics, William \& Mary, Williamsburg, VA23185}

\author{Linda Vahala}
\affil{Department of Electrical \& Computer Engineering, Old Dominion University, Norfolk, VA 12319}
\author{Abhay K. Ram}
\affil{Plasma Science and Fusion Center, MIT, Cambridge, MA 02139} 
\author{Min Soe}
\affil{Department of Mathematics  and Physical Sciences, Rogers State University, Claremore, OK 74017}

\begin{document}
\maketitle
$\bf{Abstract}$:   The effect of the thickness of the dielectric boundary layer that connects a material of refractive index $n_1$ to another of index $n_2$ is considered for the propagation of an electromagnetic pulse. For very thin boundary layer  the scattering properties of the pulse mimics that found from the Fresnel jump conditions for a plane wave - except that the transmission to incident amplitudes are augmented by a factor of $\sqrt{n_2/n_1}$.  As the boundary layer becomes thicker one finds deviations away from the Fresnel conditions and eventually one approaches WKB propagation.  However there is found a small but unusual dip in part of the transmitted pulse that persists in time.  The quantum lattice algorithm (QLA) used recovers the Maxwell equations to second order in a small parameter -- but QLA still recovers Maxwell equations when this parameter is unity.  The expansion parameter is the speed of the pulse in medium $n_1$.


\section{Introduction}
\subsection{Maxwell equations in a vacuum}

\qquad  There has been considerable interest in connecting the vacuum equations of electromagnetism with a photonic wave function [1-6].
 For example, Majorana [5]  postulated that the vacuum Maxwell equations could be rewritten in the form of a Dirac equation for the "wave function" $\mathbf{F}$ that is defined explicitly in terms of the electric and magnetic fields $\mathbf{E}$ and $\mathbf{B}$ by (under appropriate normalization)
\begin{equation}
\mathbf{F}= \mathbf{E} - i \mathbf{B}   .
\end{equation}
This function $\mathbf{F}$ had been introduced earlier [7] and is known as the Riemann-Silberstein-Weber (RSW) vector.
The vacuum Maxwell equations can then be written in terms of this RSW function
\begin{equation}
\frac{\partial \mathbf{F}}{\partial t} = i \nabla \cross \mathbf{F}    \qquad , \qquad
\nabla \cdot \mathbf{F} = 0 
\end{equation}

Using the correspondance principle
\begin{equation}
\mathcal{E} \rightarrow  i \frac{\partial}{\partial t}  \qquad , \qquad \mathbf{p} \rightarrow -i \nabla
\end{equation}
Eq. (2) takes a quasi-Dirac form
\begin{equation}
(\mathcal{E} - \boldsymbol{\alpha} \cdot \mathbf{p} ) \mathbf{F} = 0 \qquad , \qquad  
\mathbf{p} \cdot \mathbf{F} = 0
\end{equation}
where the $\boldsymbol{\alpha}$ are $3 \cross 3$ Hermitian matrices, whose specific forms are not needed here.  The first part of Eq. (4) corresponds to the curl-parts of Maxwell equations while the second part of Eq. (4) to the divergence equations.  

Khan [8] introduced a 4-spinor $\Psi$ using certain combination of the RSW vector
 \begin{equation}
\begin{aligned}
\renewcommand\arraystretch{1.6}
{\Psi}
=  
\begin{pmatrix}
- F_x + i F_y    \\
F_z \\
F_z \\
 F_x+ i F_y  
\end{pmatrix}
,
\end{aligned}
\end{equation}
and showed that the two curl and the two divergence equations of Maxwell could be written in a single 4-dimensional (4D) Dirac-like equation
\begin{equation}
\frac{\partial \Psi}{\partial t} = - \mathbf{M} \cdot \nabla \Psi 
\end{equation}
where the vector $4 \cross 4$ matrices $\mathbf{M}$ are given by the tensor products of the Pauli spin matrices
$\boldsymbol{\sigma} = (\sigma_x , \sigma_y , \sigma_z). $
with the $2 \times 2$ identity matrix $\mathbf{I_2}$:
\begin{equation}
\mathbf{M} = \boldsymbol{\sigma} \otimes \mathbf{I_2}.
\end{equation}
Indeed, the vacuum 4-spinor Maxwell Eq. (6), written out explicitly, takes the form
\begin{equation}
\frac{\partial}{\partial t}
\begin{bmatrix}
\psi_0   \\
\psi_1  \\
\psi_2   \\
\psi_3   \\
\end{bmatrix}  	
=
- \frac{\partial}{\partial x}
\begin{bmatrix}
\psi_2  \\
\psi_3  \\
\psi_0  \\
\psi_1  \\  
\end{bmatrix}
+
i \frac{\partial}{\partial y}
\begin{bmatrix}
\psi_2  \\
\psi_3 \\
-\psi_0 \\
-\psi_1  \\ 
\end{bmatrix}
-
\frac{\partial}{\partial z}
\begin{bmatrix}
\psi_0  \\
\psi_1  \\
-\psi_2  \\
-\psi_3  \\   
\end{bmatrix}
\end{equation}
while the 4-spinor Dirac equation for a massless free particle is
\begin{equation}
\frac{\partial}{\partial t}
\begin{bmatrix}
\psi_0   \\
\psi_1  \\
\psi_2   \\
\psi_3   \\
\end{bmatrix}  	
=
 \frac{\partial}{\partial x}
\begin{bmatrix}
\psi_3  \\
\psi_2  \\
\psi_1  \\
\psi_0  \\  
\end{bmatrix}
+
i \frac{\partial}{\partial y}
\begin{bmatrix}
-\psi_3  \\
\psi_2 \\
-\psi_1 \\
\psi_0  \\ 
\end{bmatrix}
+
\frac{\partial}{\partial z}
\begin{bmatrix}
\psi_2  \\
-\psi_3  \\
\psi_0  \\
-\psi_1  \\   
\end{bmatrix} .
\end{equation}

Now Yepez [9, 10] has introduced a sequence of interleaved non-commuting unitary collision and streaming operators that perturbatively recover the Dirac equation to second order accuracy.  We [11-16] then generalized these non-commuting unitary collision and streaming operators in order to generate a qubit lattice algorithm (QLA) for the solution of the vacuum Maxwell Eqs. (8).  Interestingly, these unitary QLA's
can, in principle, be readily encoded onto error-correcting qubit quantum computers to yield a quantum algorithm solving the initial value problem for Maxwell equations.

\subsection{Maxwell equations in scalar dielectric media}

\qquad  Khan[8] has also extended his matrix representation of Maxwell equations to include propagation in a scalar dielectric medium $\varepsilon(\mathbf{x})$, with refractive index $n(\mathbf{x})= \sqrt{\varepsilon(\mathbf{x})}$.  The medium inhomogeniety will now couple the two possible electromagnetic pulse polarizations so one must introduce two 4-spinors
 \begin{equation}
\begin{aligned}
\renewcommand\arraystretch{1.6}
{\Psi^{\pm} }
=  
\begin{pmatrix}
- F^{\pm}_x \pm i F^{\pm}_y    \\
F^{\pm}_z \\
F^{\pm}_z \\
 F^{\pm}_x \pm i F^{\pm}_y  
\end{pmatrix}
,
\end{aligned}
\end{equation}
with  generalized RSW vectors
 \begin{equation}
\label{R-S vector}
\mathbf{F^{\pm}} = \sqrt{\epsilon} \mathbf{E}  \pm i \frac{\mathbf{B}}{\sqrt{\mu_0}}.
\end{equation}
($\mu_0$ is the vacuum magnetic permeability).  The Maxwell equations, with no free sources, are
\begin{align*}
\nabla \bsdot \mathbf{D}  =  0 , \qquad  \nabla \bsdot \mathbf{B} =  0 
\end{align*}
\begin{equation}
\nabla \times \mathbf{E} = - \frac{\partial \mathbf{B}}{\partial t}  ,  \qquad
\nabla \times \mathbf{H} =  \frac{\partial \mathbf{D}}{\partial t} 
\end{equation}
%
with $\mathbf{D} = \varepsilon(\mathbf{x}) \mathbf{E}$ and $\mathbf{B} = \mu_0 \mathbf{H}$.  The evolution equations for the coupled 4-spinor RSW vectors [8] are
\begin{equation}
\begin{aligned}
\renewcommand\arraystretch{1.8}
\frac{\partial}{\partial t} \begin{pmatrix} \Psi^+ \\ \Psi^- \end{pmatrix}
=  
& \renewcommand\arraystretch{2.3}
 - v_{ph}
\begin{pmatrix} \mathbf{M} \bsdot \nabla - \boldsymbol{\Sigma} \bsdot 
\displaystyle{\frac{\nabla \epsilon}{4 \epsilon}}
& +i M_z \boldsymbol{\Sigma} \bsdot \displaystyle{\frac{\nabla \epsilon}{2 \epsilon}} \alpha_y \\
+i M_z \boldsymbol{\Sigma}^{*} \bsdot \displaystyle{\frac{\nabla \epsilon}{2 \epsilon}} \alpha_y &
\mathbf{M}^{*} \bsdot \nabla - \boldsymbol{\Sigma}^{*} \bsdot 
\displaystyle{\frac{\nabla \epsilon}{4 \epsilon}} \end{pmatrix}
\begin{pmatrix} \Psi^+ \\ \Psi^- \end{pmatrix} .  
\end{aligned}
\end{equation}
where $v_{ph} = (\epsilon \mu_0)^{-1/2}$ is the pulse phase velocity, with the $4 \times 4$ matrices
\begin{equation}
\boldsymbol{\alpha} = 
\begin{pmatrix} 
0 & \boldsymbol{\sigma}  \\
\boldsymbol{\sigma} & 0 \\
\end{pmatrix},
\ \ \ 
\boldsymbol{\Sigma} = 
\begin{pmatrix} 
\boldsymbol{\sigma} & 0  \\
0 & \boldsymbol{\sigma} \\
\end{pmatrix}.
\end{equation}
Now the time evolution of the 8-spinor $(\Psi^+ , \Psi^-)$ is no longer fully Hermitian [8] - but several of the matrices (which depend on $\nabla \epsilon$) are anti-Hermitian.  In particular, for 1D pulse propagation in the z-direction, the 8-spinor system, Eqs. (13), reduces to 
\begin{align*}
\label{R_S InHomox1}
\frac{\partial}{\partial t}
\begin{bmatrix}
\psi_0   \\
\psi_1  \\
\psi_2   \\
\psi_3   \\
\end{bmatrix}  	
= - \frac{1}{n(z)} 
\frac{\partial}{\partial z}
\begin{bmatrix}
\psi_0  \\
\psi_1  \\
-\psi_2  \\
-\psi_3  \\ 
\end{bmatrix}
+ \frac{n^\prime(z)}{2n^2(z)}
\begin{bmatrix}
\psi_0 - \psi_7  \\
-\psi_1 - \psi_6  \\
\psi_2 + \psi_5  \\
-\psi_3 + \psi_4  \\  
\end{bmatrix}
\end{align*}
\begin{equation}
\label{R_S InHomox1}
\frac{\partial}{\partial t}
\begin{bmatrix}
\psi_4   \\
\psi_5  \\
\psi_6   \\
\psi_7   \\
\end{bmatrix}  	
= - \frac{1}{n(z)} 
\frac{\partial}{\partial z}
\begin{bmatrix}
\psi_4  \\
\psi_5  \\
-\psi_6  \\
-\psi_7  \\ 
\end{bmatrix}
+ \frac{n^\prime (z)}{2n^2(z)}
\begin{bmatrix}
\psi_4 - \psi_3  \\
-\psi_5 - \psi_2  \\
\psi_6 + \psi_1  \\
-\psi_7 + \psi_0  \\  
\end{bmatrix}.
\end{equation}

To develop a QLA for the 1D z-propagation of a pulse in a dielectric medium we need to determine the specific collision and streaming operators  required so that Eqs. (15) are recovered perturbatively to 2nd order accuracy.  
The collision operator must couple two qubits at a given lattice site so that we have quantum entanglement, and then that entanglement is spread throughout the lattice by the streaming.  If we look at the time evolution of any of the 8-spinor $\psi_i$ in Eqs. (15), one sees that it is coupled to the z-derivative $\partial \psi_i /\partial z$ of the same spinor component $\psi_i$.
Hence for a QLA for z-propagation we are forced into a $16$-qubit representation.  This coupling of the same spinor component in $\partial /\partial t$ with  $\partial /\partial z$ can be traced to the diagonal property of the Pauli $\sigma_z$ matrix.  This does not occur in the QLA for x- or y-propagation of the 8-spinor since different spinor components of  $\partial \psi_i/\partial t$ couple with  $\partial \psi_j/\partial x$ and  $\partial \psi_j/\partial y$ with $j \neq i$ [11-14].  Thus for x- and y-propagation we can directly work with an 8-qubit set that is just the 8-spinor set of Khan's Maxwell representation.  
To develop QLA for 2D and 3D simulations, the interleaved sequence of collide-streams in each of the Cartesian directions will act consecutively on the chosen qubit basis.  It is thus more convenient to work with a 16-qubit basis by simply generalizing the 8-spinor set to the 16-qubit space.

\subsection{QLA for 1D Propagation in the z-direction}
\qquad  We connect the 16-qubit field $Q = (\overline{q}_0 .... \overline{q}_{15})^T$  to the 8-spinor field $(\psi_0  ...  \psi_7)^T$   by 
\begin{align*}
\psi_0 = \overline{q}_0+\overline{q}_2  ,  \psi_1 = \overline{q}_1+\overline{q}_3, \psi_2 = \overline{q}_4+\overline{q}_6, \psi_3 = \overline{q}_5+\overline{q}_7,
\end{align*}
\begin{equation}
\psi_4 = \overline{q}_8+\overline{q}_{10}  ,  \psi_5 = \overline{q}_9+\overline{q}_{11}, \psi_6 = \overline{q}_{12}+\overline{q}_{14}, \psi_7 = \overline{q}_{13}+\overline{q}_{15}.
\end{equation}
The QLA on the 16-qubit basis consists of an interleaved sequence of collision ($C$) and streaming ($S$) operators
\begin{equation}
\begin{aligned}
\label{U_X}
U &= \overline{S}_{-} C  \overline{S}_{+} C^{\dagger} \bsdot \: S_{+} C  S_{-} C^{\dagger}, \\
 \overline{U} &= S_{+} C^{\dagger}  S_{-} C \bsdot \: \overline{S}_{-} C^{\dagger}  \overline{S}_{+} C
\end{aligned}
\end{equation}
where $C^\dagger$ is the adjoint of $C$.  The operator $S_+$ streams the
8-qubit subset $[\overline{q}_0, \overline{q}_1, \overline{q}_4, \overline{q}_5, \overline{q}_8, \overline{q}_9, \overline{q}_{12}, \overline{q}_{13}]$
one lattice unit in the positive z-direction.  Notice that these qubits are each the first qubits on the right hand sides of Eqs. (16).
$S_-$ streams these 8 qubits one lattice unit in the negative z-direction.  Similarly for the streaming operator $\overline{S}$ for the other 8 qubit which are the second qubits on the right hand sides of Eqs. (16).

The spatial and time derivatives on the 8-spinor field are determined from $\overline{U} . U  Q(t)$.
In particular, an appropriate $16 \times 16$ unitary collision operator $C$  couples the qubits [$\overline{q}_0 - \overline{q}_2$,  $\overline{q}_1 - \overline{q}_3$],   [$\overline{q}_4 - \overline{q}_{6}$, $\overline{q}_5 - \overline{q}_7$],  [$\overline{q}_8 - \overline{q}_{10}$,  $\overline{q}_9 - \overline{q}_{11}$],  [$\overline{q}_{12} - \overline{q}_{14}$, $\overline{q}_{13} - \overline{q}_{15}$]
and has $4 \cross 4$ block diagonal structure
\begin{equation}
\label{C4x4}
C(\theta) = 
\begin{bmatrix}
C_4 (\theta) & 0 & 0 & 0   \\
0  &  C_4 (\theta)^{T}  & 0 & 0  \\  
0  & 0 & C_4 (\theta)  &  0 \\
 0  &  0 & 0 & C_4 (\theta)^{T} \\ 
\end{bmatrix},
\end{equation}
with the $4 \cross 4$ submatrix

\begin{equation}
\label{Csub4x4}
C_4 (\theta) = 
\begin{bmatrix}
\cos \theta & 0 & \sin \theta & 0   \\
0  &  \cos \theta  & 0 & \sin \theta  \\  
-\sin \theta  & 0 &  \cos \theta  &  0 \\
 0  &  -\sin \theta & 0 & \cos \theta  \\ 
\end{bmatrix},
\end{equation} 
and collision angle $\theta$ 
\begin{equation}
\theta = \frac{\epsilon}{4 n(z)}
\end{equation}
for some small perturbation parameter $\epsilon$.  This sequence of collide-stream operators will generate, in the continuum limit, the required first term $\partial / \partial z$ in Eqs. (15) to $O(\epsilon^2)$.

To recover the two $n^\prime (z)$ terms in the 8-spinor Eqs. (15) one introduces a particular potential collision operator for each term.  Since this first $n^\prime (z)$ term has exactly the same couplings to the spinor-components
as the $\partial /\partial t$ and $\partial /\partial z$ terms,
the appropriate QLA potential collision operator will again have this $4 \cross 4$ diagonal block structure
\begin{equation}
\label{C4x4}
\overline{P}_{1Z} (\gamma) = 
\begin{bmatrix}
P_4 (\gamma) & 0 & 0 & 0   \\
0  &  P_4 (\gamma) & 0 & 0  \\  
0  & 0 & P_4 (\gamma) &  0 \\
 0  &  0 & 0 & P_4 (\gamma) \\ 
\end{bmatrix},
\end{equation} 
with
\begin{equation}
\label{Csub4x4}
P_4 (\gamma) = 
\begin{bmatrix}
\cos \gamma & 0 &- \sin \gamma & 0   \\
0  &  \cos \gamma  & 0 & - \sin \gamma \\  
- \sin \gamma & 0 & \cos \gamma  &  0 \\
 0  &  - \sin \gamma & 0 &\cos \gamma \\ 
\end{bmatrix},
\end{equation}
for some collisional angle $\gamma$.  The second  potential collision operator has diagonal structure in its two $8 \cross 8$ matrices
\begin{equation}
\label{Csub4x4}
\overline{P}_{2Z} (\gamma) = 
\begin{bmatrix}
P_{81} (\gamma) & P_{82} (\gamma)   \\
P_{82} (\gamma) & P_{81} (\gamma)  \\  
\end{bmatrix},
\end{equation}
where
\begin{equation}
\label{PV81 8x8}
P_{81} (\gamma) = 
  \begin{bmatrix}
  \cos \gamma & 0 & 0 & 0 & 0 & 0 & 0 & 0   \\
  0  &  \cos \gamma  &  0  & 0  &  0  &  0  &  0  &  0 \\
  0  &  0  &  \cos \gamma  &  0  & 0 & 0 & 0 & 0   \\
  0  &  0  &  0  &  \cos \gamma  &  0  &  0  &  0  &  0 \\ 
   0 & 0 & 0 & 0  &  \cos \gamma & 0 &  0 & 0 \\
   0 & 0 & 0 & 0  &  0  &  \cos \gamma &  0  &  0 \\
    0 & 0 & 0 & 0  & 0  &  0  &  \cos \gamma  &  0 \\
     0 & 0 & 0 & 0  &  0  & 0  &  0  &  \cos\gamma  \\
   \end{bmatrix}.
\end{equation} 
and
\begin{equation}
\label{PV82 8x8}
P_{82} (\gamma) = 
  \begin{bmatrix}
 0 & 0 & 0 & 0 & 0 & 0 & 0 & - \sin \gamma   \\
  0  & 0  &  0  & 0  &  0  &  0  &  - \sin \gamma  &  0 \\
  0  &  0  & 0 &  0  & 0 & - \sin \gamma & 0 & 0   \\
  0  &  0  &  0  & 0 &  - \sin \gamma  &  0  &  0  &  0 \\ 
   0 & 0 & 0 &  \sin \gamma   &  0 & 0 &  0 & 0 \\
   0 & 0 &  \sin \gamma  & 0  &  0  &  0 &  0  &  0 \\
    0 &  \sin \gamma  & 0 & 0  & 0  &  0  &  0  &  0 \\
      \sin \gamma  & 0 & 0 & 0  &  0  & 0  &  0  &  0  \\
   \end{bmatrix}.
\end{equation} 
To recover the 8-spinor Eqs. (15) to $O(\epsilon^2)$, one chooses
\begin{equation}
\gamma = \varepsilon^2 \frac{n^\prime (z)}{2 n(z)^2}.
\end{equation}
The final QLA that is used for the simulations is 
\begin{equation}
Q(t+\delta t) = \overline{P}_{2Z}(\gamma). \overline{P}_{1Z}(\gamma). \overline{U} . U. Q(t)
\end{equation}

\subsection{Initial conditions for the 16 qubit QLA}

We are given the initial electromagnetic field components of the pulse.  From the RSW transformation, Eq. (10), we then have the initial 8-spinor $(\psi_0 , .... \psi_7)$.  The  initial condition for the 16-qubit $Q(t=0)$  is then chosen:
\begin{align*}
 \overline{q}_0 = \overline{q}_2 = \psi_0 /2  , \overline{q}_1 = \overline{q}_3 = \psi_1 /2 , \overline{q}_4 = \overline{q}_6 = \psi_2 /2 , \overline{q}_5 = \overline{q}_7 = \psi_3 /2 ,
\end{align*}
\begin{equation}
 \overline{q}_8 = \overline{q}_{10} = \psi_4 /2 ,  \overline{q}_9 = \overline{q}_{11} = \psi_5 /2 , \overline{q}_{12} = \overline{q}_{14} = \psi_6 /2 , \overline{q}_{13} = \overline{q}_{15} = \psi_7 /2 .
\end{equation}

\section{Effect of Dielectric Boundary Layer Thickness on Reflection and Transmission of a 1D Pulse}

Our earlier QLA simulations [11, 12] of an electromagnetic pulse propagating  from one dielectric (with refractive index $n_1$) to another (with refractive index $n_2$) at normal incidence assumed the boundary layer thickness connecting the two media, $\Delta_{BL}$, was very much less than the width of the incident pulse $\Delta_{pulse}$:  $\Delta_{BL} << \Delta_{pulse}$.  In this case we found that the ratio of the reflected to initial field amplitudes were just those derived from the Fresnel equations for a $\emph{boundary value problem}$ of a $\emph{plane wave}$ normally incident onto a dielectric discontinuity [17].  However, for pulse propagation as an initial value problem, the transmitted to incident field amplitude is augmented from the plane wave Fresnel results by a factor $\sqrt{n_2/n_1}$ :
\begin{equation}
\frac{E_{refl}}{E_{inc}	} = \frac{n_1 -n_2}{n_1 + n_2} \qquad ,\quad  \frac{E_{trans}}{E_{inc}	} = \frac{2 n_1}{n_1 + n_2} \sqrt{\frac{n_2}{n_1}} .
\end{equation}
As the QLA simulations repeated stamped out Eq. (29) for many different simulations with various choices of $n_1$ and $n_2$ and different pulse geometries, we [12] developed a theory for Gaussian pulses that validates Eq. (29).  A back-of-the-envelope argument also shows this:  
now the reflected pulse preserves all the reflected Fresnel plane wave characteristics.  Thus they must have the energy.  Since the total energy of the scattering is conserved, then we must have the same conservation of transmitted energy.
However, the speed of propagation of the transmitted pulse and its width are reduced by a factor of $n_2/n_1$.  Hence the amplitude of the transmitted pulse must increase by a factor of $\sqrt{n_2/n_1}$.  

\subsection{Case 1 :  $\Delta_{BL} << \Delta_{pulse}$ with $\epsilon = 0.3$}
We consider the case of $n_1=1$ (for $z<4990$) and $n_2=2$ (for $z>5010$), with the boundary layer to pulse width (see Fig. 1) of $\Delta_{BL} / \Delta_{pulse} = 0.1$.  The vertical dashed line in Fig. 1 is the mid point of the boundary layer.
\begin{figure}[!h!] \ 
\begin{center}
\includegraphics[width=3.2in]{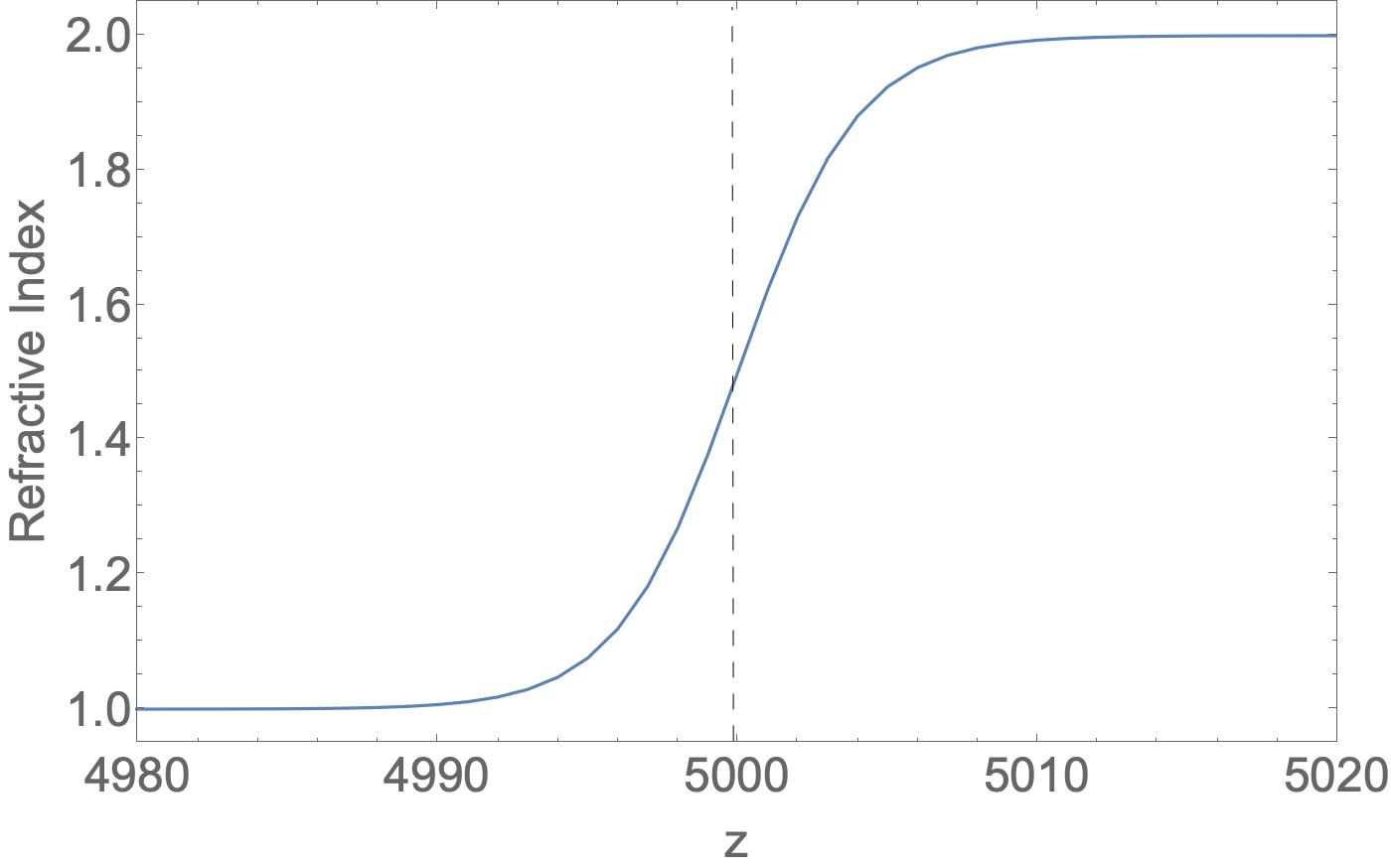}
\includegraphics[width=3.2in]{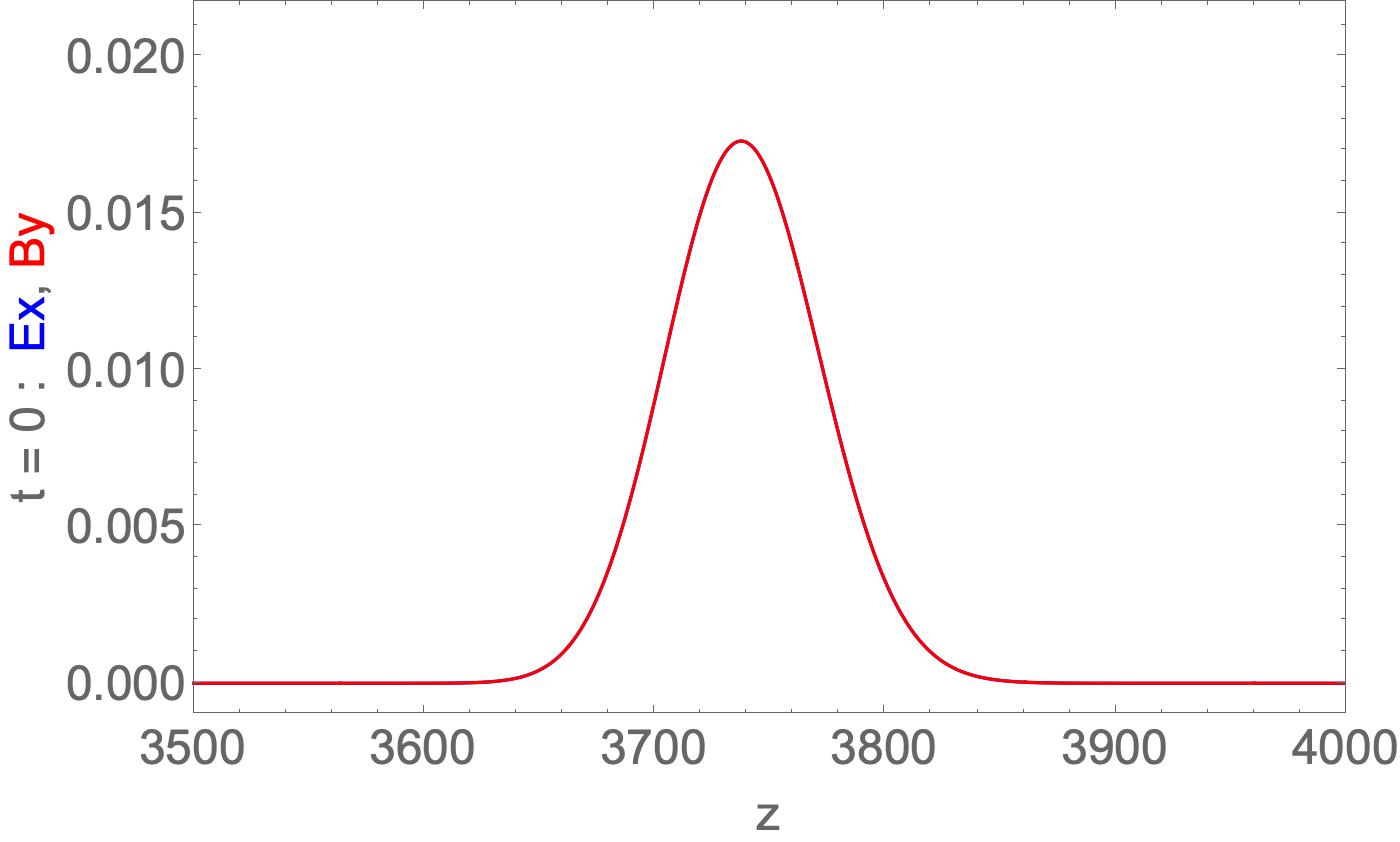}
(a)  dielectric boundary layer  \quad  \quad    (b)  initial $E_x , B_y$ profile
\caption{(a) A dielectric boundary layer connecting medium $n_1=1$, for $z < 4990$ to medium $n_2=2$ for $z > 5010$.  The vertical dashed line indicates the midpoint of the boundary layer, (b)  The initial vacuum fields (in our units) $E_x = B_y$ with $\Delta_{pulse} = 200$.  Initially $B_y = n_1 E_x$ and the two field profiles overlay each other.  $E_x$ - blue, $B_y$ - red.
}
\end{center}
\end{figure} 
The perturbation parameter $\epsilon$, introduced into the QLA collision angles Eq. (20) and (26), is chosen to be $\epsilon=0.3$.
In the QLA simulation units, the time  increment $\delta t = 1$, Eq. (27).

As the pulse propagates into the boundary layer, part of it will be transmitted and part of it will  be reflected.  Away from the boundary layer, asynptotically, the transmitted amplitudes have $B_y = (n_2 /n_1) E_x$, while the reflected amplitudes have $B_y = - E_x > 0$. (Fig. 2).  From QLA simulations, we find
\begin{equation}
\frac{E_{refl}}{E_{inc}	} =- 0.32\qquad ,\quad  \frac{E_{trans}}{E_{inc}	} = 0.94  \qquad .
\end{equation}
This is in excellent agreement with Eq. (29), while the Fresnel plane wave boundary conditions would yield $ E_{trans}/E_{inc}	= 0.67$.
\begin{figure}[!h!] \ 
\begin{center}
\includegraphics[width=3.2in]{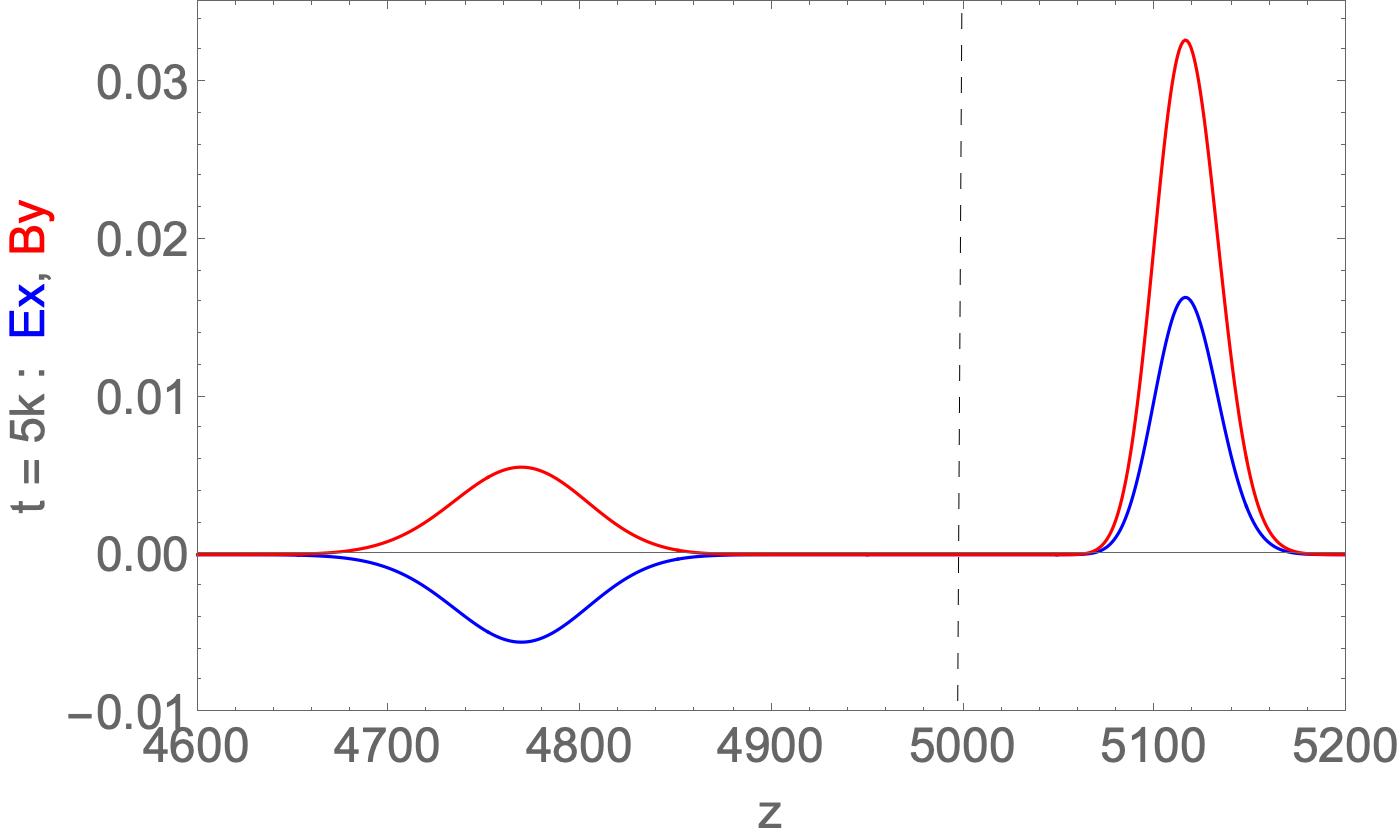}
\includegraphics[width=3.2in]{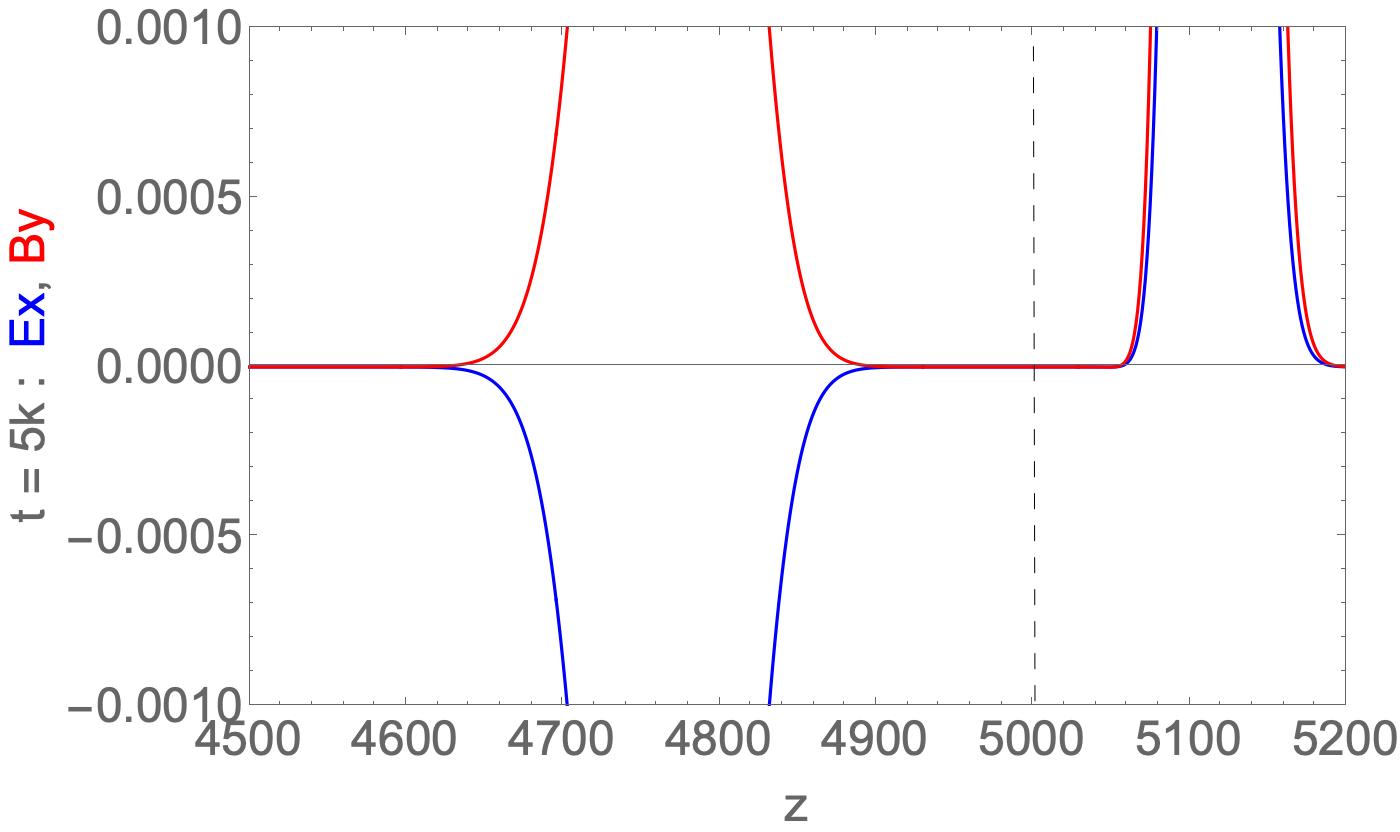}
(a)  reflected and transmitted fields  \quad  \quad    (b) blowup of the profiles
\caption{(a) The transmitted and reflected fields with $\Delta_{BL} << \Delta_{pulse}$ after 5000 time iterations. (i.e., at t = 5k).  (b)  A blowup showing the Gaussian nature of the reflected and transmitted profiles.  Note there is no disturbance between the reflected and transmitted profiles.  $E_x$ - blue, $B_y$ - red.
}
\end{center}
\end{figure} 
Note that the width of the reflected pulse is equal to that of the incident pulse, but twice that of the transmitted pulse $\Delta_{refl} = 2 \Delta_{trans}$, and the transmitted pulse travels at half the speed of the reflected pulse.

\subsection{Case 2 :  $\Delta_{BL} \approx \Delta_{pulse}$ with $\epsilon = 0.3$}
When the boundary layer thickness is on the order of the pulse width, Fig. 3a,
both the transmitted and reflected field amplitudes are significantly affected.
 In Fig. 3b, at time $t = 4k$ much of the pulse is within this boundary layer $4900 < z < 5100$.  There is developing characteristics of the $n_2 = 2 $ region as the magnetic field amplitude is basically twice that of the electric field amplitude for $z > 5000$ while part of the pulse for $z < 5000$ is showing the characteristics of the $n_1 = 1$ region with a significant range having $E_x < 0$.
\begin{figure}[!h!] \ 
\begin{center}
\includegraphics[width=3.2in]{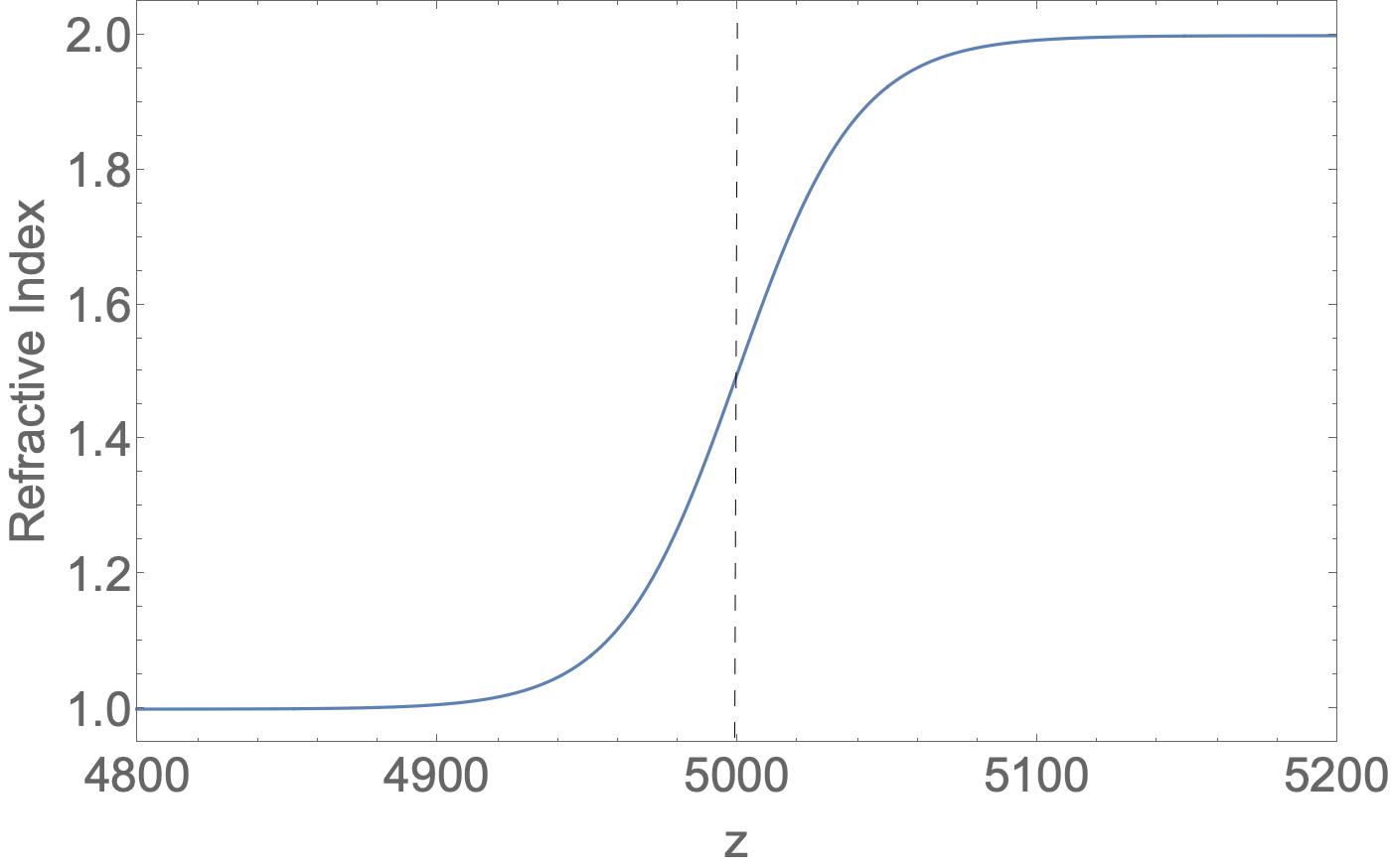}
\includegraphics[width=3.2in]{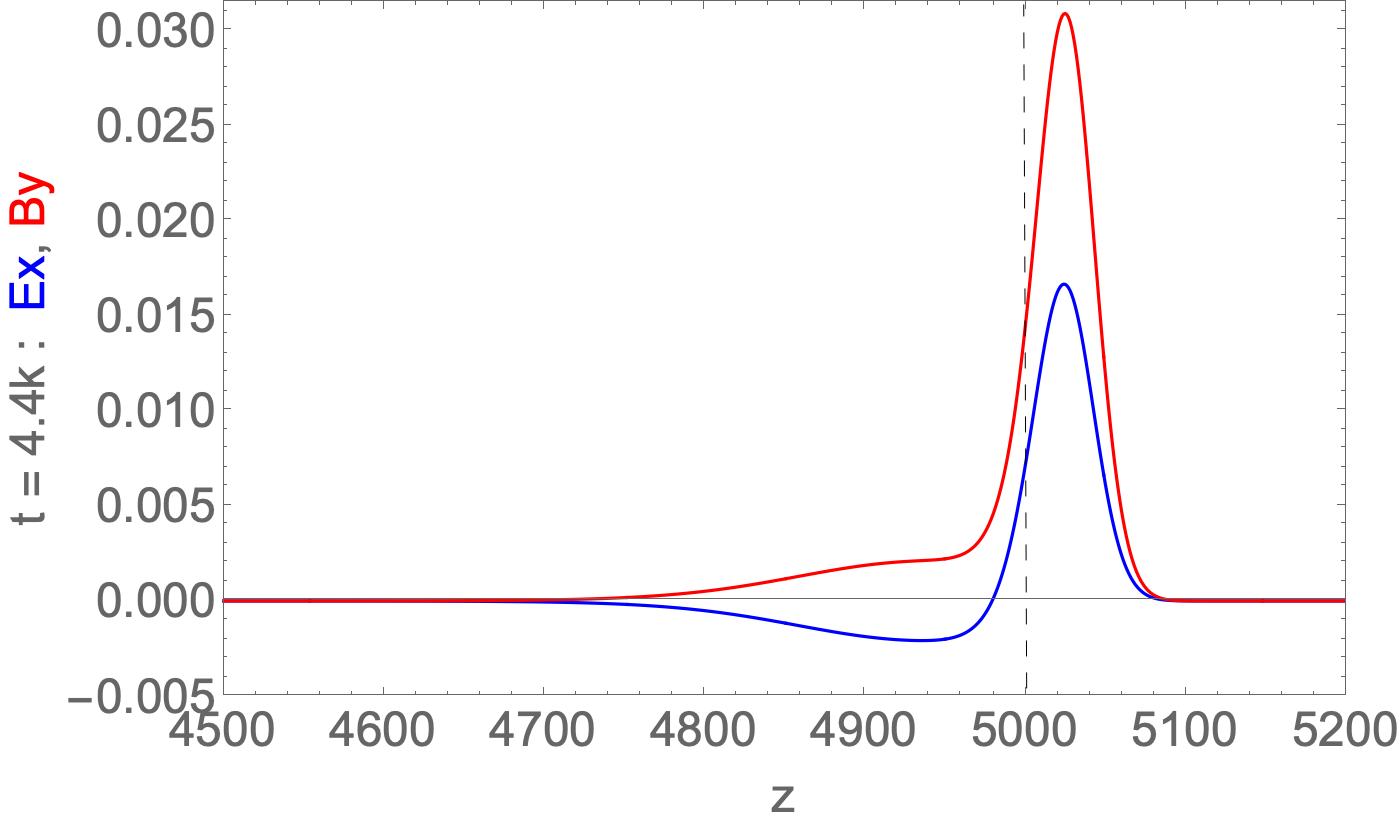}
(a) thicker dielectric boundary layer \quad  \quad    (b) pulse overlapping the boundary layer
\caption{(a) The dielectric boundary layer now extends $4900 < z < 5100$, with $\Delta_{BL} \approx \Delta_{pulse}$  (b)  At $t = 4k$, the pulse is overlapping the boundary layer.  The basic 2:1 ratio of the transmitted magnetic to transmitted electric field is being established for $z > 5000$.  $E_x$ -  blue, $B_y$ - red.  In the reflected pulse, a significant portion already has $E_x < 0$ since $n_1 < n_2$.
}
\end{center}
\end{figure} 

By $t = 5k$, the reflected and transmitted pulses are approaching their quasi-asymptotic state.  The reflected pulse has a lower amplitude and greater width than that predicted by the Fresnel plane wave conditions, Fig. 4a.
 Of some interest is the trailing edge of the transmitted pulse which exhibits a slight dip with both $B_y < 0$ and $E_x< 0$, but with $B_y/E_x \approx 2$, Fig. 4b.
\begin{figure}[!h!] \ 
\begin{center}
\includegraphics[width=3.2in]{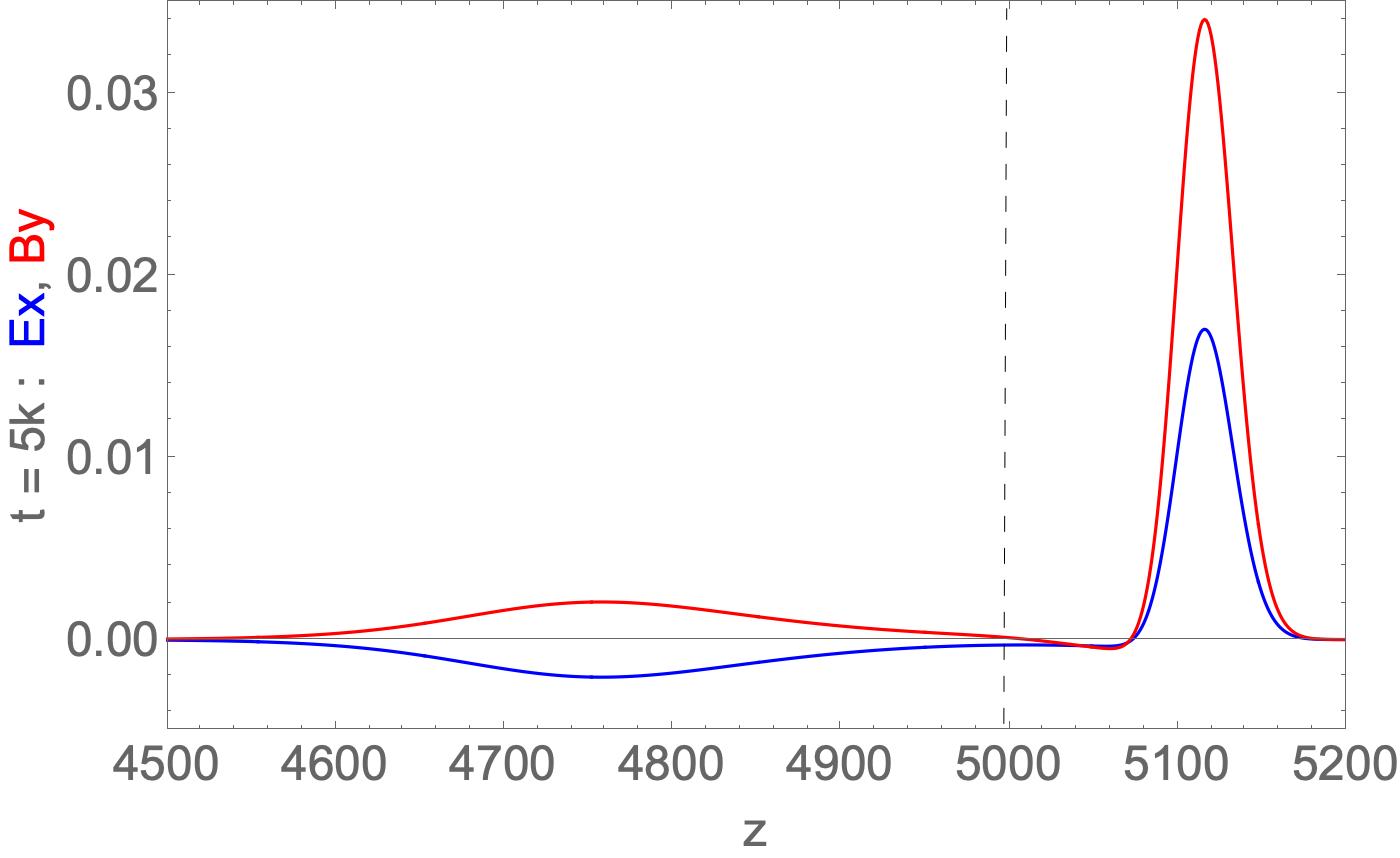}
\includegraphics[width=3.2in]{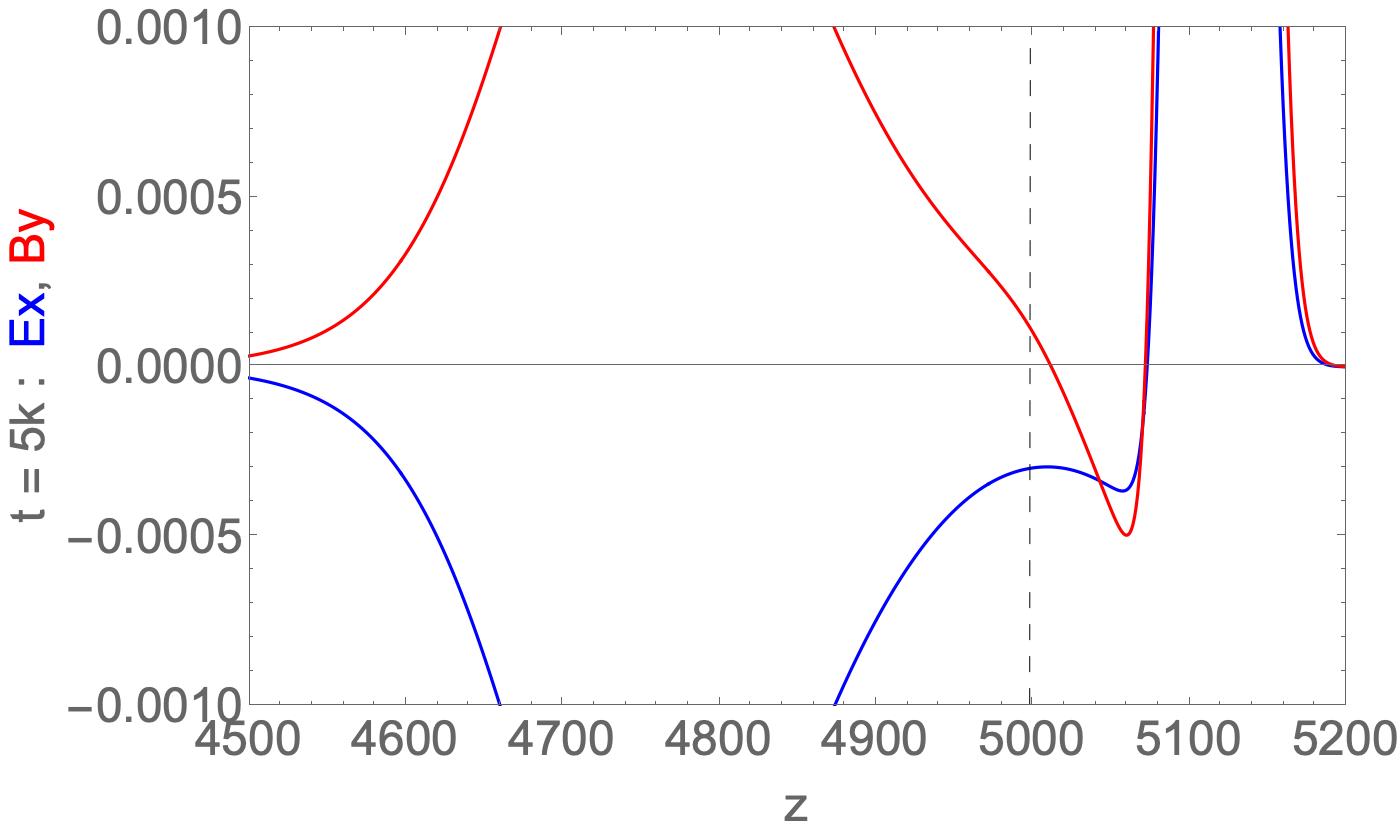}
(a) pulse at $t = 5k$ \quad  \quad    (b) blowup of pulse at $t=5k$
\caption{(a) The transmitted and reflected fields when $\Delta_{BL} \approx \Delta_{pulse}$.  The reflected amplitude is almost a factor of 3 lower than the reflected Fresnel plane wave solution and the reflected pulse is significantly broader.  (b)  A blowup showing asymmetry in both the reflected and transmitted profiles.  There is a small region in $z > 5000$ in which the transmitted pulse exhibits the unusual behavior of $B_z < 0$ and $E_z < 0$.  $E_x$ -  blue, $B_y$ - red. 
}
\end{center}
\end{figure} 

In Fig. 5 we see the asymptotic profiles for the reflected and transmitted pulses.  $E_{refl}/E_{inc} = -0.12$ while the Fresnel plane wave solution yields $-0.33$.
\begin{figure}[!h!] \ 
\begin{center}
\includegraphics[width=3.2in]{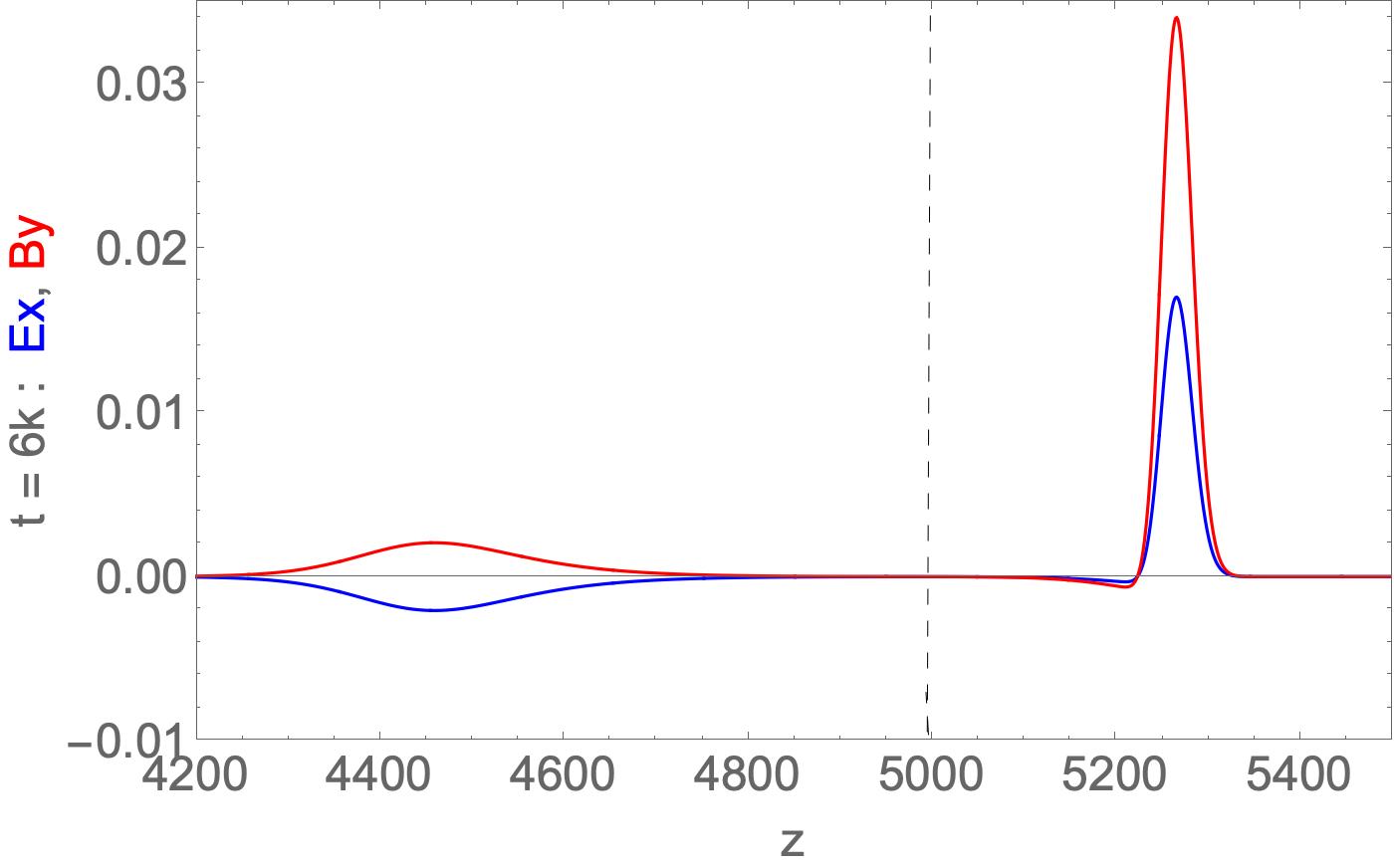}
\includegraphics[width=3.2in]{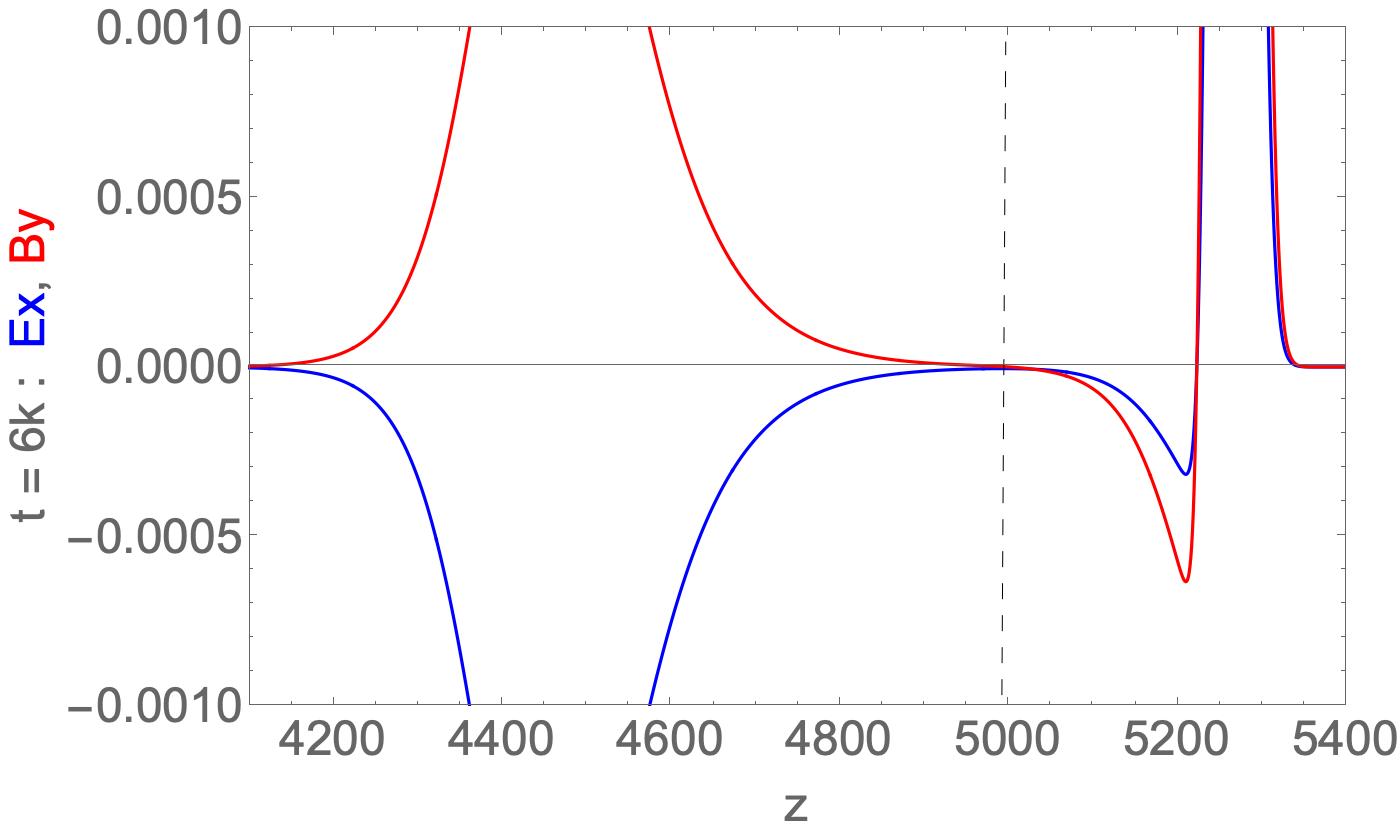}
(a) pulse at time $t = 6k$ \quad  \quad    (b) Blowup of the pulse at $t = 6k$.
\caption{(a) The transmitted and reflected fields with $\Delta_{BL} << \Delta_{pulse}$.  $E_x$ - blue, $B_y$ - red. (b)  A blowup showing the asymmetric nature of the reflected and transmitted profiles.  At the back end of the transmitted pulse one finds a small region in which both $B_y < 0$ and $E_x < 0$ and with $B_y \approx 2 E_x$.
}
\end{center}
\end{figure} 
Around $z \approx 5210$ one finds a section of the transmitted pulse that has $B_y < 0 , E_x < 0$ with the magnetic to electric field peaks being in the 2:1 ratio, as might be expected in the $n_2=2$ dielectric region.  This feature is asymptotically stable and will continue traveling as a part of the transmitted pulse.  It is not present when $\Delta_{BL} << \Delta_{pulse}$, Fig. 2b.

\subsection{Case 3:   $\Delta_{BL} >> \Delta_{pulse}$ with $\epsilon = 0.3$ }
We now consider the case when the boundary layer thickness $\Delta_{BL} \approx 2000$, and $\Delta_{pulse} \approx 200$, Fig 6a.
\begin{figure}[!h!] \ 
\begin{center}
\includegraphics[width=3.2in]{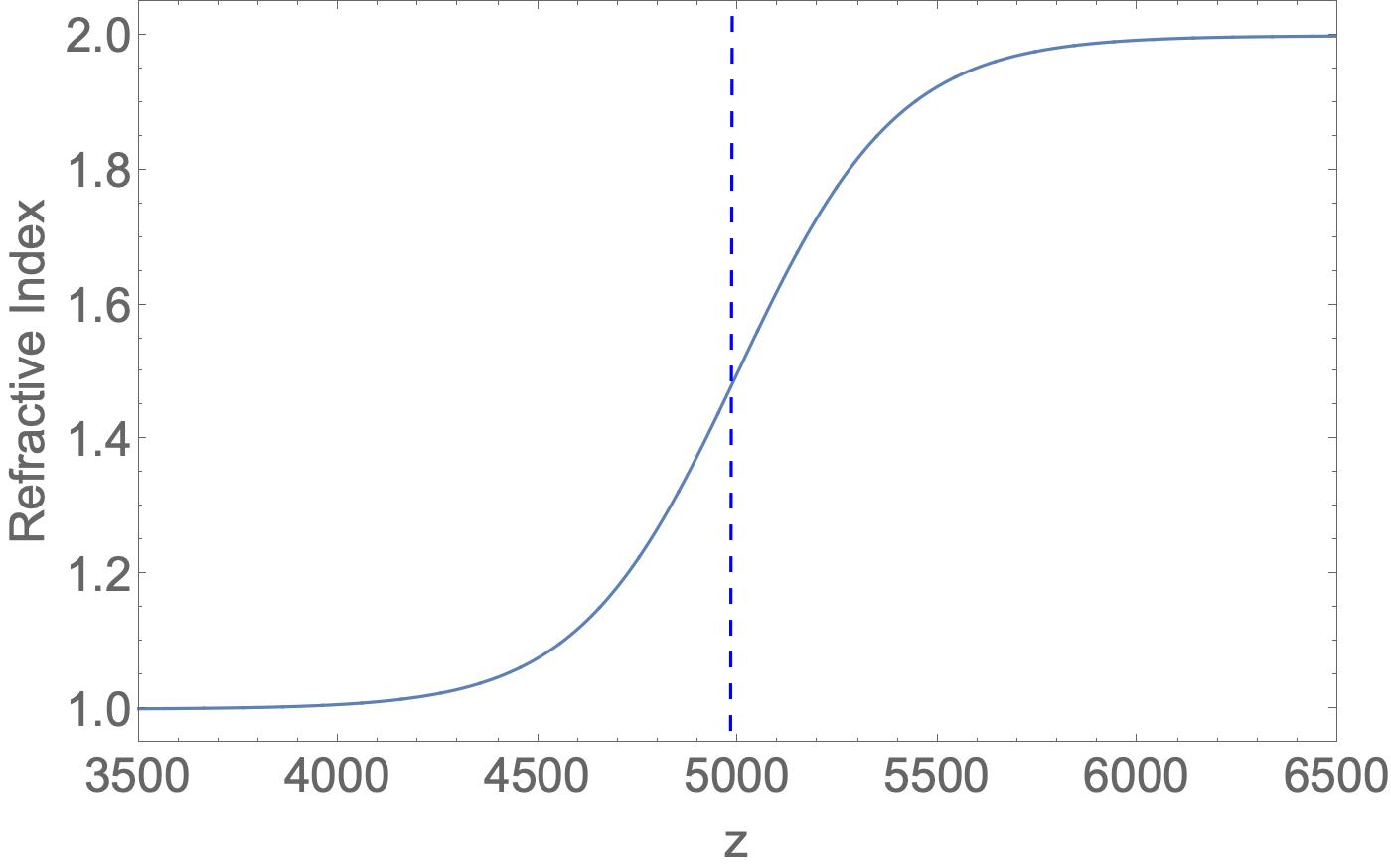}
\includegraphics[width=3.2in]{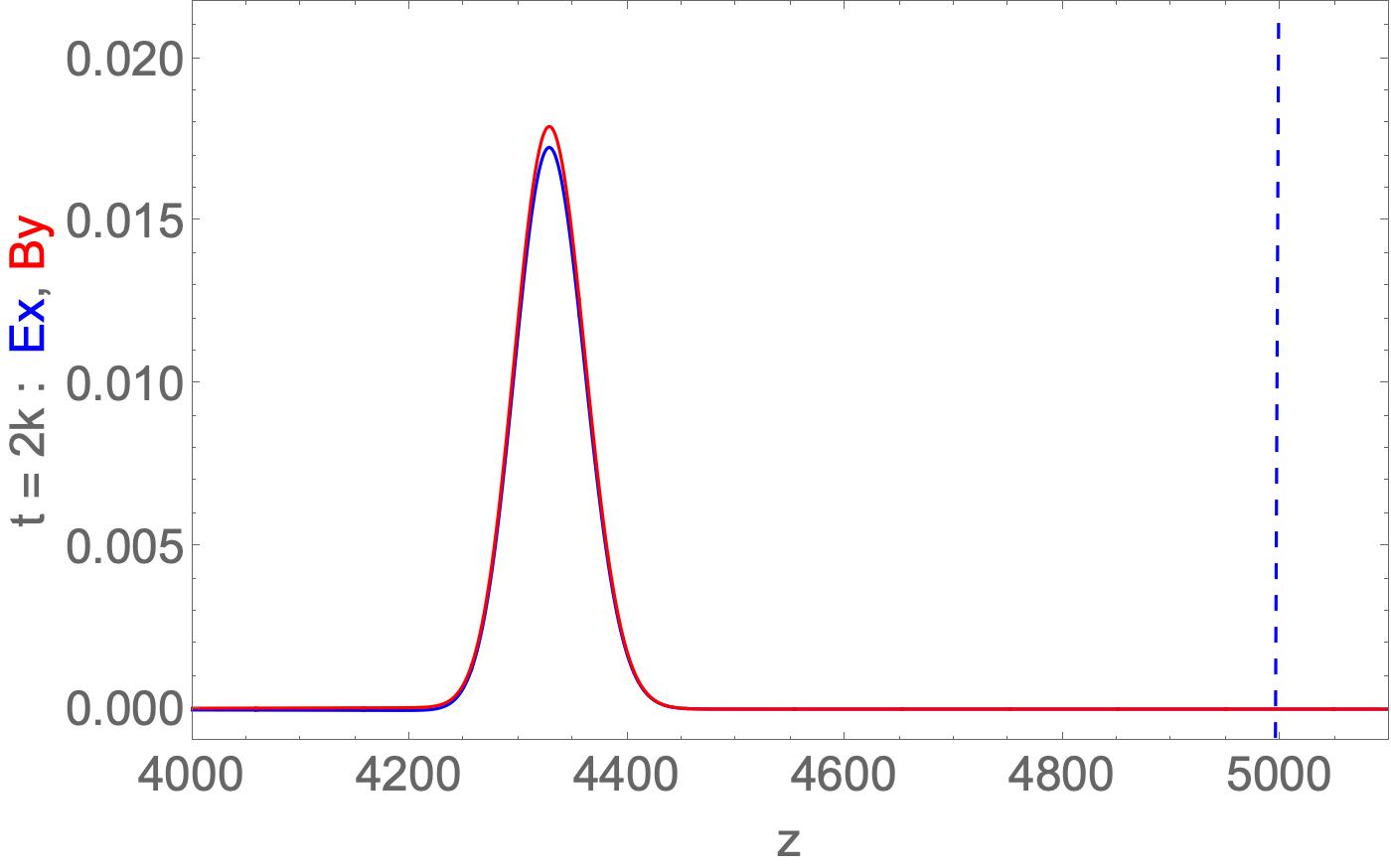}
(a)  refractive index with $\Delta_{BL} >> \Delta_{pulse}$.  \qquad  \qquad \qquad  (b) Pulse at $t = 2k$
\caption{(a) The refractive index for the case when $\Delta_{BL} \approx 2000$ and $\Delta_{pulse} \approx 200 $.
(b)  By $2k$ iterations the pulse has moved into a slightly non-vacuum dielectric state, as seen by the amplitude peak in $B_y$ ebing greater than that in $E_x$.   $E_x$ -  blue, $B_y$ - red. 
}
\end{center}
\end{figure} 
As the pulse propagates, Fig. 6b-8b, the peak amplitude ratio of $B_y/E_x$ basically scales as the value of the local refractive index at that point.  In essence,  the time evolution of the pulse is similar to WKB .  There is a very low order reflected pulse  - but its peaks are 2 orders of magnitude less than the transmitted pulse (Fig. 8b).
\begin{figure}[!h!] \ 
\begin{center}
\includegraphics[width=3.2in]{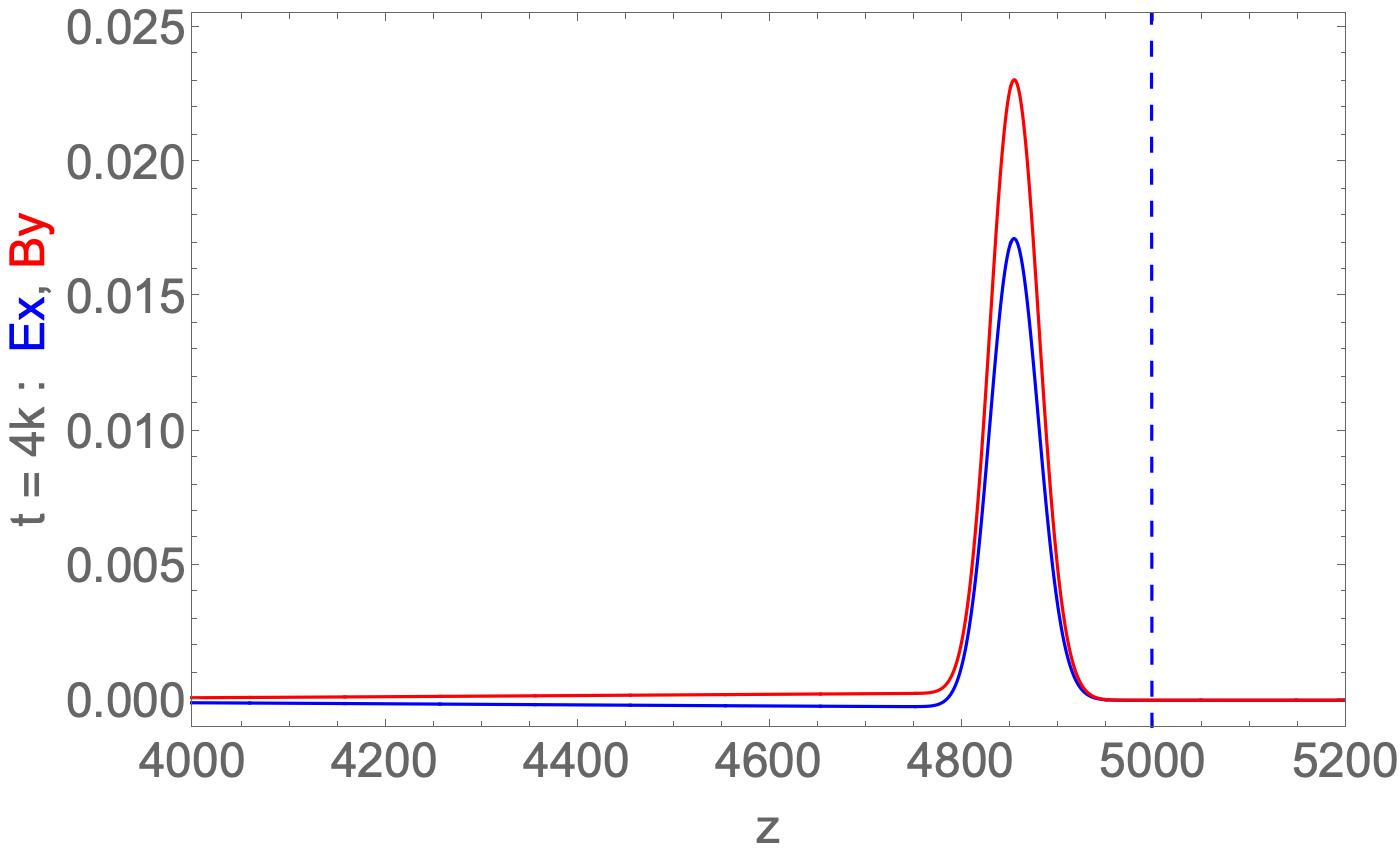}
\includegraphics[width=3.2in]{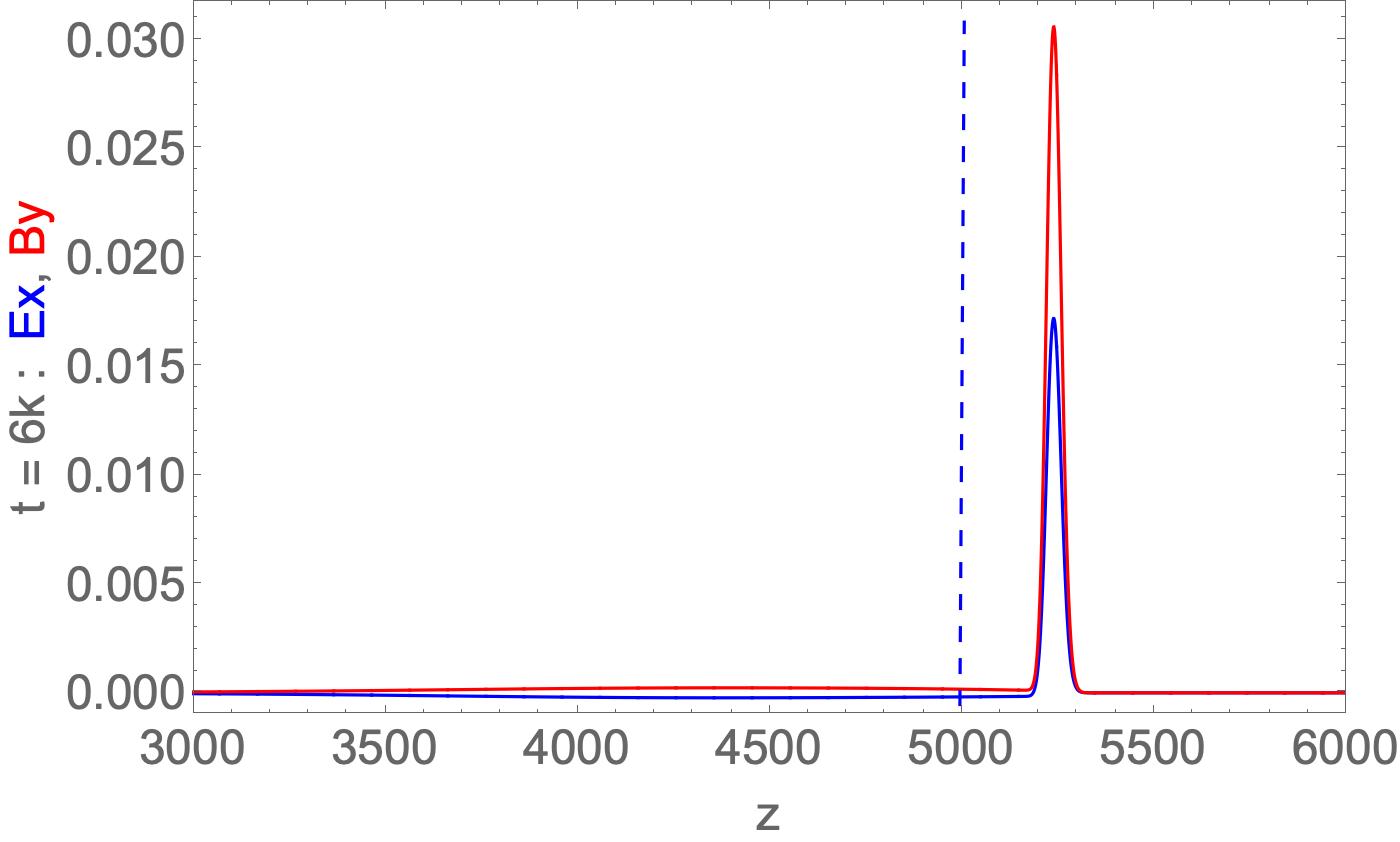}
(a) slowly varying refractive index   (b) Pulse at $t = 3k$
\caption{(a) The refractive index for the case when $\Delta_{BL} >> \Delta_{pulse}$.
(b)  By $2k$ iterations the pulse has moved into a slightly non-vacuum state, as seen by the amplitude peak in $B_y$ ebbing greater than that in $E_x$.   $E_x$ -  blue, $B_y$ - red. 
}
\end{center}
\end{figure} 
\begin{figure}[!h!] \ 
\begin{center}
\includegraphics[width=3.2in]{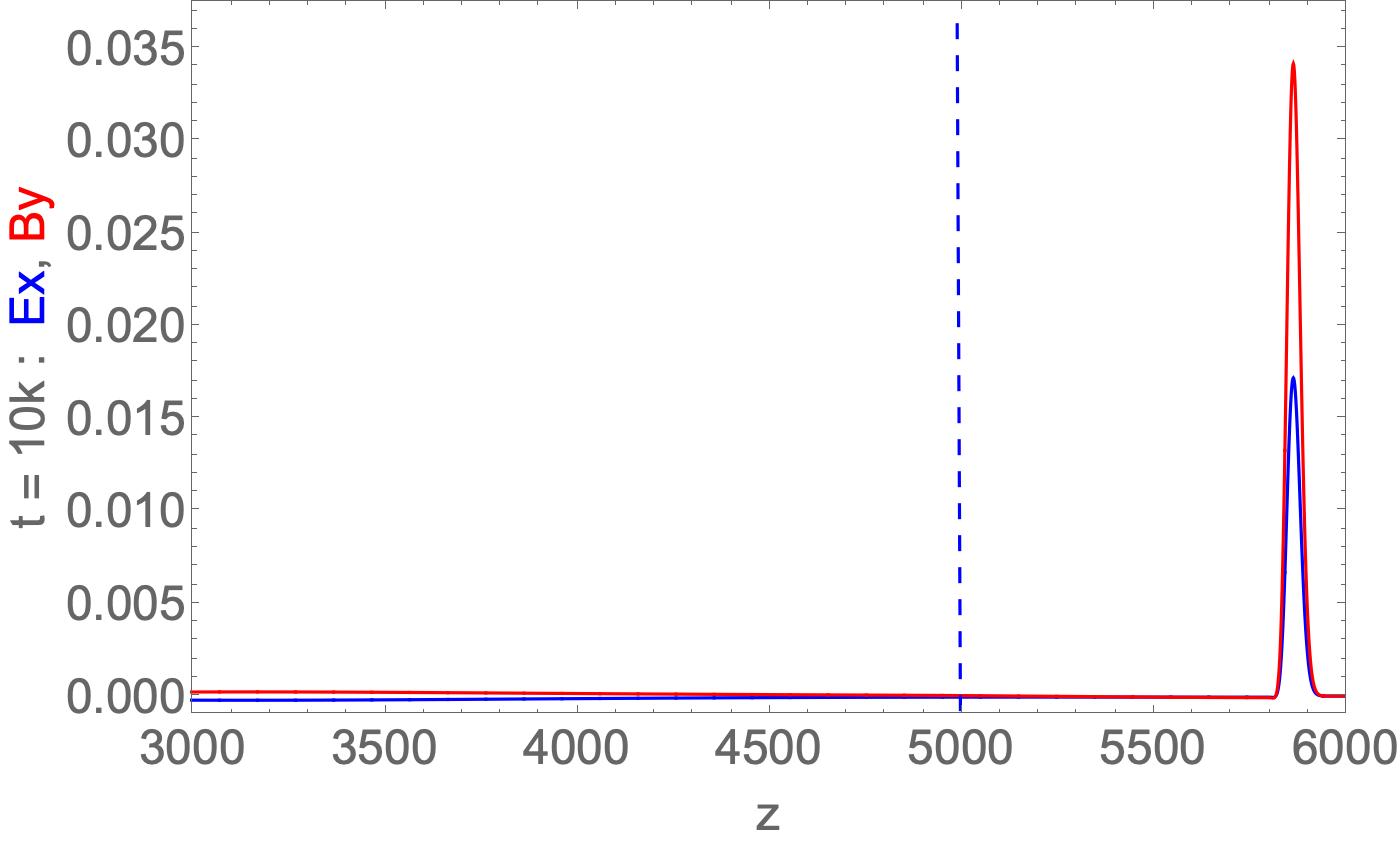}
\includegraphics[width=3.2in]{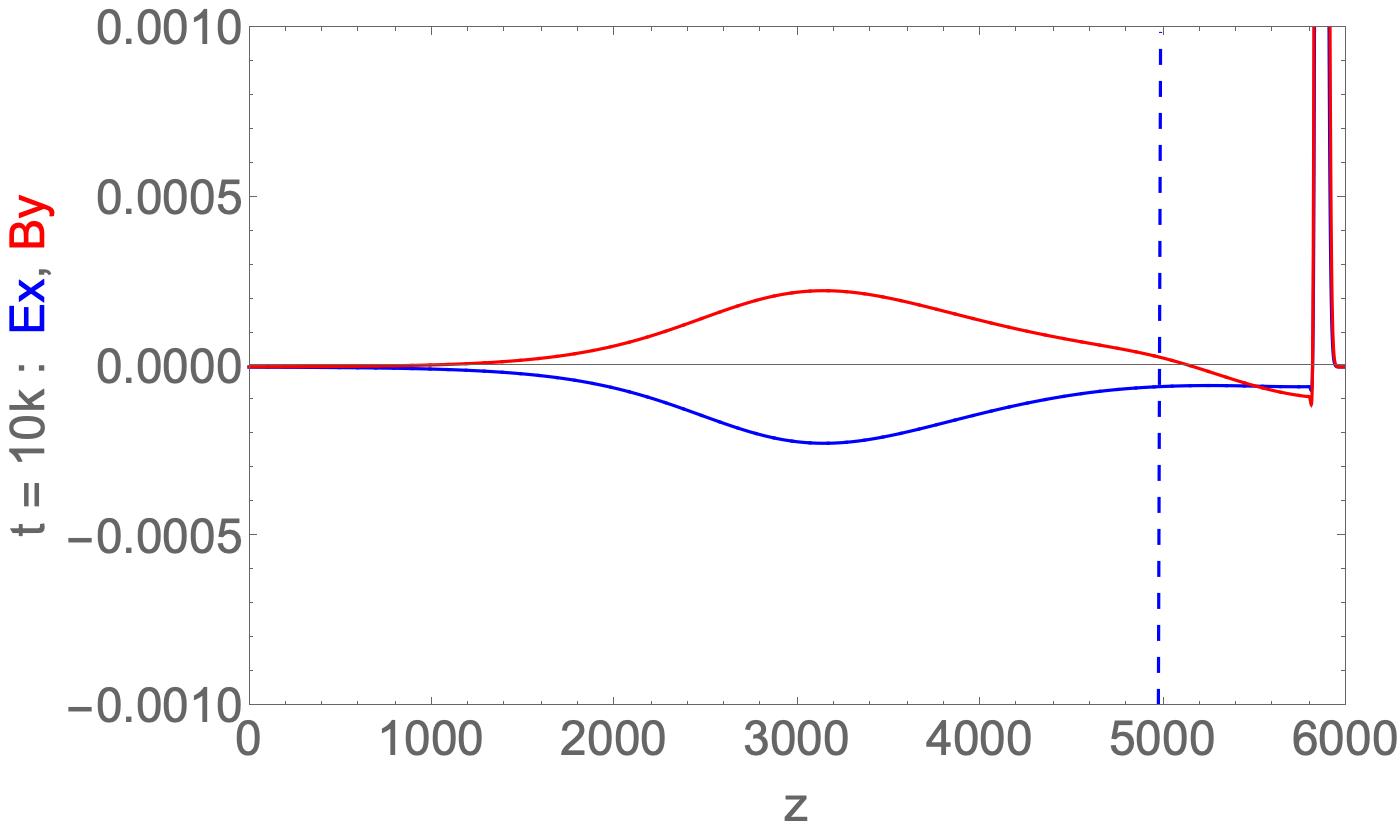}
(a) slowly varying refractive index   (b) Pulse at $t = 3k$
\caption{(a) The refractive index for the case when $\Delta_{BL} >> \Delta_{pulse}$.
(b)  By $2k$ iterations the pulse has moved into a slightly non-vacuum state, as seen by the amplitude peak in $B_y$ ebing greater than that in $E_x$.   $E_x$ -  blue, $B_y$ - red. 
}
\end{center}
\end{figure} 

\subsection{Case 4 :  $\Delta_{BL} \approx \Delta_{pulse}$ with $\epsilon = 1.0$ }
\qquad  QLA is a perturbation theory, based on the parameter $\epsilon$ introduced into the various collision operators, Eqs. (20) and (26).   Moreover, QLA recovers the Maxwell equations only to errors $O(\epsilon^2)$.  In our earlier QLA for solitons [18]  the perturbation parameter was related to the amplitude of the wave function $\psi_{NLS}$..  Since the evolution equations for solitons/quantum vortices involve a $|\psi_{NLS}|^2$ nonlinearity, we [18] found that QLA simulations strongly deviated away from the exact nonlinear soliton dynamics if $\epsilon$ was chosen too high - typically $\epsilon \approx 0.45$.  These deviations took the form of background turbulent noise along with distorted to disintegrating soliton shapes.

Here we are developing a QLA for the linear Maxwell equations and find that the basic Maxwell equations are still being modeled even up to $\epsilon = 1.0$.  The essential physics is retained (Fig. 9) but there is a slight deviation in the pulse propagation speed.  Indeed $\epsilon$ is nothing but the pulse speed in medium 1.  For $\epsilon=0.15$, in 20k time steps the pulse peak location is delayed by 7 lattice units.   For $\epsilon=0.3$ , by $t= 10k$, the pulse peak is delayed by 16 lattice units.  This delay increases for increasing $\epsilon$:  for $\epsilon = 0.6$, the peak is off by 45 lattice units, and by $125$ lattice units for $\epsilon=1.0$.
\begin{figure}[!h!] \ 
\begin{center}
\includegraphics[width=3.2in]{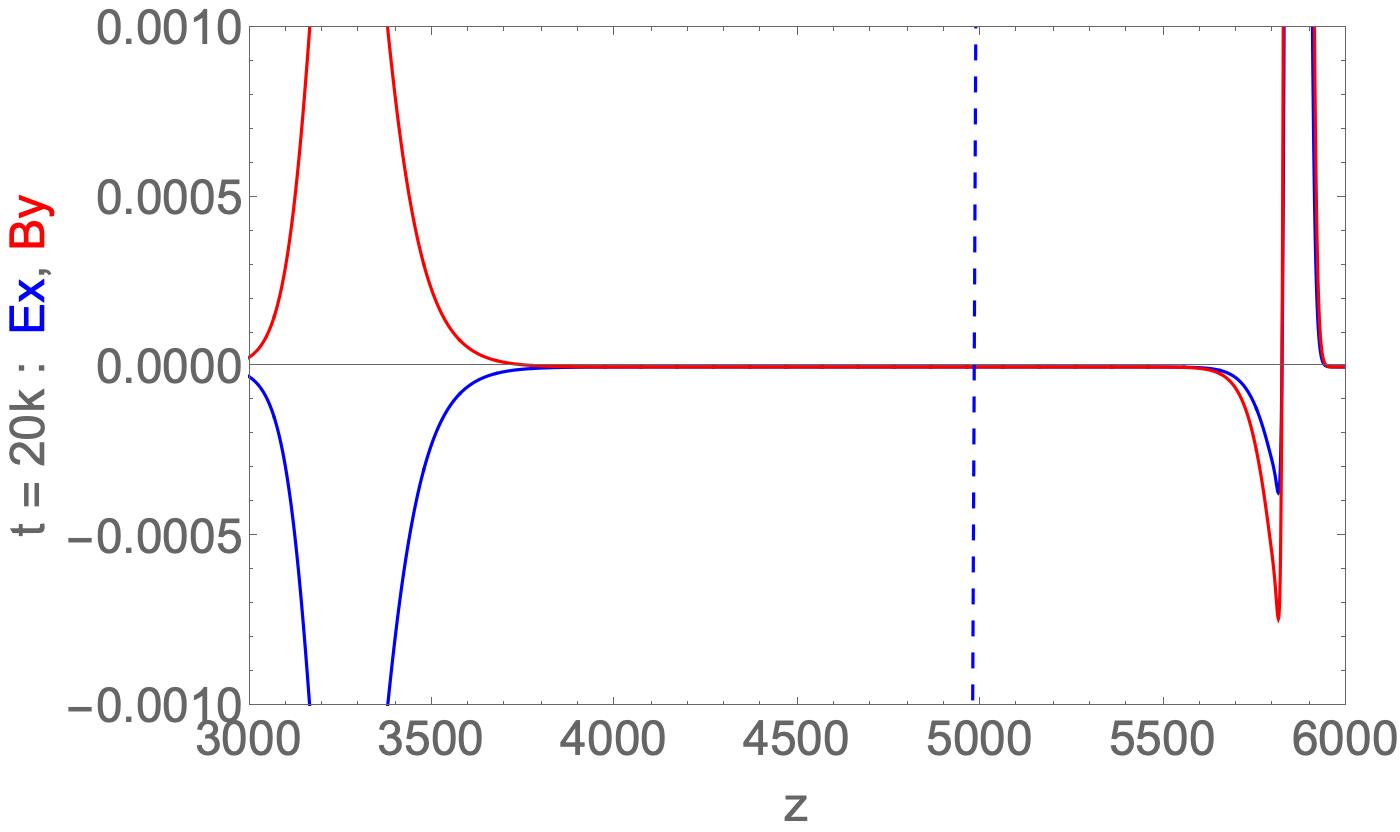}
\includegraphics[width=3.2in]{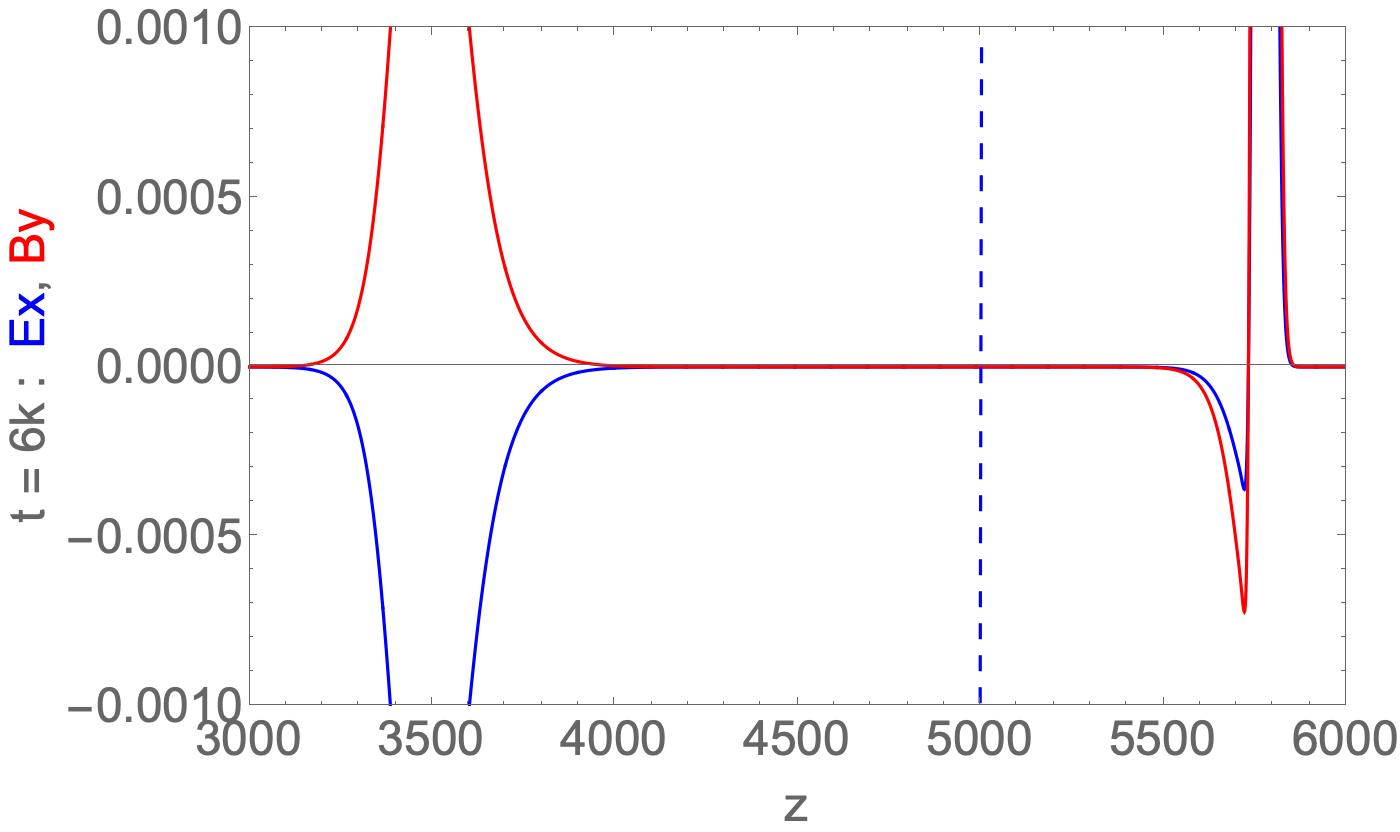}
(a) $\epsilon = 0.3, t = 20k$ \qquad  \qquad \qquad  \qquad  (b) $\epsilon = 1.0,  t = 6k$
\caption{(a) A blowup of the reflection-transmission pulses for the case $\Delta_{BL} \approx \Delta_{pulse}$ at t = 20k and with $\epsilon = 0.3$
(b) The corresponding pulses but now for $\epsilon = 1.0$ and at $t = 6K$.  For this higher $\epsilon$ there is a slight decrease in the QLA pulse propagation speed.   $E_x$ -  blue, $B_y$ - red. 
}
\end{center}
\end{figure} 

\section{Summary and Conclusions}
\qquad   In this paper we have examined the effect of the boundary layer thickness on the 1D normal propagation of a pulse in a dielectric medium.  
For very sharp boundary layers connecting the two dielectric regions, our initial value QLA simulations give reflected and transmitted electromagnetic fields in agreement with the boundary value Fresnel plane wave conditions - except that the ratio of the transmitted to incident field amplitudes is augmented by a factor $\sqrt(n_2/n_1)$.  
As the thickness of the boundary layer, $\Delta_{BL}$, increases the reflected pulse becomes more and more modified.  It becomes smaller in amplitude as well as increasing in width till we approach the WKB-like solution when 
$\Delta_{BL} >> \Delta_{pause}$.  A somewhat unsuspected feature was seen in the trailing edge of the transmitted pulse.  We noticed a region in the $n_2$ dielectric in which both $B_y < 0$ and $E_x < 0$.  This "dip" is stable and propagates away undistorted from the boundary layer along with the rest of the transmitted pulse.  We have also performed a QLA run in which $n_1 = 2 > n_2= 1$.  In region $n_1$ we initially have $B_y = 2 E_x$.  The transmitted pulse now has $B_y = E_x$, but again with a "dip in the transmitted fields with $B_y = E_x < 0.$.  The Poynting flux has been determined for each of the QLA runs, and we find energy is conserved with normalized variations on the order of $1.35 \times 10^{-3}$. 

Our interest in the effects of the boundary layer thickness on scattering stems from some of our 2D QLA simulations [15] where we have considered a plane 1D electromagnetic pulse scattering from a small dielectric cylinder.  For $\Delta_{BL} << |Delta_{pulse}$ we have found quite complicated structures being emitted from the dielectric cylinder due to the internal bouncing from the cylinder walls.  For example, in Fig. 10, we have a 1D plane pulse propagating to along the x- axis.  In the middle of the plane is a 2D dielectric cylinder of internal refractive index $n_2=3$ whose diameter $ >> \Delta_{pulse}$.  The pulse speed within the dielectric cylinder is reduced by a factor of $3$ over the rest of the plane propagation as it continues along the axis.  However, within the cylinder the pulse will bounce and then undergo reflection and transmission at the dielectric cylinder boundaries.  This accounts for the multiple circular wave fronts for the electric field being emitted at various times in the QLA simulation.  Fig. 10 is a snapshot of $E_y(x,z)$ at time t = $20k$.  The 1D pulse can be seen  as a dark thin strip to the right of the dielectric cylinder.
\begin{figure}[!h!] \ 
\begin{center}
\includegraphics[width=6.2in]{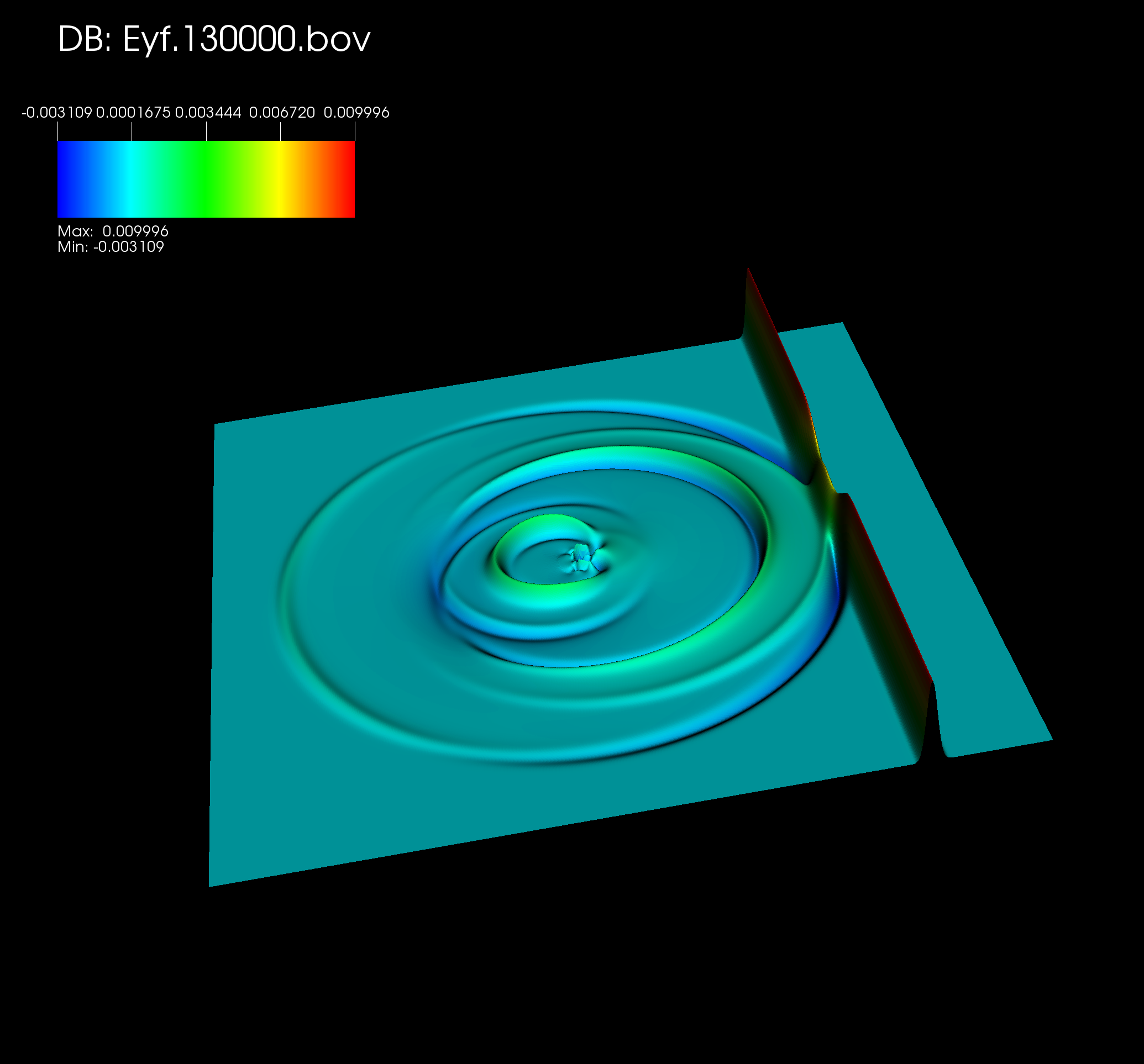}
\caption{The scattering from a dielectric cylinder centered at  $x=L/2, z=L/2$ of a plane 1D pulse propagating along the x-axis.  L is the number of lattice sites in each direction.  Plotted is the electric field $E_y(x,z,t)$ at time $t = 20k$.  The multiple circular ring disturbances arise from the internal scattering and then transmission out of the dielectric cylinder regime.
}
\end{center}
\end{figure} 
\section{Acknowledgments}
This research was partially supported by Department of Energy grants DE-SC0021647, DE-FG02-91ER-54109, DE-SC0021651, DE-SC0021857, and DE-SC0021653.  The 2D QLA simulation was performed on $CORI$ of the National Energy Research Scientific Computing Center (NERSC), A U.S. Department of Energy Office of Science User Facility located at Lawrence Berkeley National Laboratory, operated under Contract No. DE-AC02-05CH11231.  The 1D QLA simulations were performed on a laptop.


\section{References}
\quad [1] O. Laporte $\&$ E. G. Uhlenbeck, “Application of spinor analysis to the Maxwell and Dirac    equations”, Phys. Rev. 37, 1380-1397 (1931)

[2]  J. R. Oppenheimer, “Note on light quanta and the electromagnetic field”, Phys. Rev. 38, 725-746 (1931).

[3] E. Moses, “Solutions of Maxwell’s equations in terms of a spinor notation:  the direct and inverse problems”,  Phys. Rev. 113, 1670-1679 (1959)

[4]  I. Bialynicki-Birula, “Photon Wave Function” , in Progress in Optics, Vol. 34, pp. 248-294, ed. E. Wolf (North-Holland, 1996).

[5]  E. Majorana (unpublished notes) : quoted after R. Mignani, E. Recami $\&$ M. Baldo, “About a Diraclike Equation for the Photon, according to Ettore Majorana.”, Lett.  Nuovo Cimento, 11, 568-572 (1974).

[6]  S. Esposito, “Covariant Majorana Formulation of Electrodynamics”, Foundations of Physics 28, 231-244 (1998)

[7]  L. Silberstein, Annalen der Physik 22, 579 (1907); 24, 783 (1907)

[8]  S. A. Khan, “Maxwell Optics:  I.  An exact matrix representation of the Maxwell equations in a medium”,  Physica Scripta 71, 440-442 (2005);  also arXiv: 0205083v1 (2002)

[9]  Y. Yepez, "An efficient and accurate quantum algorithm for the Dirac equation", arXiv: 0210093 (2002).

[10]  J. Yepez, "Relativistic Path Integral as a Lattice-Based Quantum Algorithm", Quant. Info. Proc. \textbf{5}, 471-509 (2005)

[11]  G. Vahala, L. Vahala, M. Soe $\&$ A. K. Ram,  "Unitary Quantum Lattice Simulations for Maxwell Equations in Vacuum and in Dielectric Media" , J. Plasma Phys. 86m 905860518 (2020)

[12]  A. K. Ram, G. Vahala, L. Vahala, $\&$ M. Soe. "Reflection and transmission of electromagnetic pulses at a planar dielectric interface - theory and quantum lattice simulations", AIP Advances 11, 105116 (2021)

[13]  G. Vahala, L. Vahala, M. Soe $\&$ A. K. Ram, “One- and two-dimensional quantum lattice algorithms for Maxwell equations in inhomogeneous scalar dielectric media I:  theory”
Rad. Effects Defects in Solids, 176, 49-63 (2021)

[14]  G. Vahala, M. Soe, L. Vahala $\&$ A. K. Ram, “One- and two-dimensional quantum lattice algorithms for Maxwell equations in inhomogeneous scalar dielectric media II:  Simulations”
Rad. Effects Defects in Solids, 176, 64-72 (2021)

[15]  G. Vahala, M. Soe, L. Vahala $\&$ A. K. Ram,  “Two Dimensional Electromagnetic Scattering from Dielectric Objects using Quantum Lattice Algorithm”, arXiv:2110.05480

[16]  G. Vahala, J. Hawthorne, L. Vahala, A. K. Ram $\&$ M. Soe, “Quantum lattice representation for the curl equations of Maxwell equations”, arXiv:2111.09745

[17]  J. D.  Jackson,  “Classical Electrodynamics”, 3rd Ed., (Wiley, New York, 1998)

[18] A. Oganesov, G. Vahala, L. Vahala, J. Yepez $\&$ M. Soe, “Benchmarking the Dirac-generated unitary lattice qubit collision-stream algorithm for 1D vector Manakov soliton collisions”,  Computers Math. with Applic.  72, 386 (2016)

\end{document}